\documentclass[
twocolappendix, twocolumn]{aastex631}
\usepackage{amsmath}

\usepackage{comment}

\usepackage{soul}

\begin{document}

\title{Constraining the z $\sim$ 1 Initial Mass Function with {\it HST} and {\it JWST} Lensed Stars in MACS J0416.1-2403}

\author[0000-0002-4490-7304]{Sung Kei Li}
\email{keihk98@connect.hku.hk}
\affiliation{Department of Physics, The University of Hong Kong, Pokfulam Road, Hong Kong}

\author[0000-0001-9065-3926]{Jose M. Diego}
\affiliation{IFCA, Instituto de F\'isica de Cantabria (UC-CSIC), Av. de Los Castros s/n, 39005 Santander, Spain}

\author[0000-0002-7876-4321]{Ashish K. Meena}
\affiliation{Department of Physics, Ben-Gurion University of the Negev, PO Box 653, Be’er-Sheva 8410501, Israel}

\author[0000-0003-4220-2404]{Jeremy Lim}
\affiliation{Department of Physics, The University of Hong Kong, Pokfulam Road, Hong Kong}


\author[0000-0002-5899-3936]{Leo W. H. Fung}
\affiliation{Institute for Computational Cosmology \& Centre for Extragalactic Astronomy, Durham University, Stockton Rd, Durham DH1 3LE, UK}

\author[0009-0000-0660-1219]{Arsen Levitskiy}
\affiliation{Centre for Astrophysics and Supercomputing, Swinburne University of Technology, Hawthorn, VIC 3122, Australia}

\author[0000-0001-6985-2939]{James Nianias}
\affiliation{Department of Physics, The University of Hong Kong, Pokfulam Road, Hong Kong}

\author[0000-0003-0942-817X]{Jose M. Palencia}
\affiliation{IFCA, Instituto de F\'isica de Cantabria (UC-CSIC), Av. de Los Castros s/n, 39005 Santander, Spain}

\author[0000-0002-1681-0767]{Hayley Williams}
\affiliation{Minnesota Institute for Astrophysics, 
University of Minnesota, 116 Church St. SE, Minneapolis, MN 55455, USA}

 \author[0000-0002-3783-4629]{Jiashuo Zhang}
\affiliation{Department of Physics, The University of Hong Kong, Pokfulam Road, Hong Kong}

\author[0000-0003-1276-1248]{Alfred Amruth}
\affiliation{Department of Physics, The University of Hong Kong, Pokfulam Road, Hong Kong}

\author[0000-0002-8785-8979]{Thomas J. Broadhurst}
\affiliation{Department of Theoretical Physics, University of Basque Country UPV/EHU, Bilbao, Spain}
\affiliation{Ikerbasque, Basque Foundation for Science, Bilbao, Spain}
\affiliation{Donostia International Physics Center, Paseo Manuel de Lardizabal, 4, San Sebasti\'an, 20018, Spain}

\author[0000-0003-1060-0723]{WenLei Chen}
\affiliation{Department of Physics, Oklahoma State University, 145 Physical Sciences Bldg, Stillwater, OK 74078, USA}

\author[0000-0003-3460-0103]{Alexei V. Filippenko}
\affiliation{Department of Astronomy, University of California, Berkeley, CA 94720-3411, USA}

\author[0000-0003-3142-997X]{Patrick L. Kelly}
\affiliation{Minnesota Institute for Astrophysics, 
University of Minnesota, 116 Church St. SE, Minneapolis, MN 55455, USA}

\author[0000-0002-6610-2048]{Anton M. Koekemoer}
\affiliation{Space Telescope Science Institute, 3700 San Martin Drive,
Baltimore, MD 21218, USA}

\author[0000-0002-4693-0700]{Derek Perera}
\affiliation{Minnesota Institute for Astrophysics, 
University of Minnesota, 116 Church St. SE, Minneapolis, MN 55455, USA}

\author[0000-0001-7957-6202]{Bangzheng Sun}
\affiliation{Department of Physics and Astronomy, University of Missouri, Columbia, MO 65211, USA}

\author[0000-0002-6039-8706]{Liliya L. R. Williams}
\affiliation{Minnesota Institute for Astrophysics, 
University of Minnesota, 116 Church St. SE, Minneapolis, MN 55455, USA}

\author[0000-0001-8156-6281]{Rogier A. Windhorst}
\affiliation{School of Earth and Space Exploration, Arizona State University,
Tempe, AZ 85287-6004, USA}

\author[0000-0001-7592-7714]{Haojin Yan}
\affiliation{Department of Physics and Astronomy, University of Missouri, Columbia, MO 65211, USA}

\author[0000-0002-0350-4488]{Adi Zitrin}
\affiliation{Department of Physics, Ben-Gurion University of the Negev, PO Box 653, Be’er-Sheva 8410501, Israel}




\begin{abstract}

Our understanding of galaxy properties and evolution is contingent on knowing the initial mass function (IMF), and yet to date, the IMF is constrained only to local galaxies. Individual stars are now becoming routinely detected at cosmological distances, where luminous stars such as supergiants in background galaxies strongly lensed by galaxy clusters are temporarily further magnified by huge factors (up to $10^{4}$) by intracluster stars, thus being detected as transients. The detection rate of these events depends on the abundance of luminous stars in the background galaxy and is thus sensitive to the IMF and the star-formation history (SFH), especially for the blue supergiants detected as transients in the rest-frame ultraviolet/optical filters. As a proof of concept, we use simple SFH and IMF models constrained by spectral energy distributions (SEDs) to see how well we can predict the {\it HST} and {\it JWST} transient detection rate in a lensed arc dubbed ``Spock'' ($z = 1.0054$). We find that demanding a simultaneous fit of the SED and the transient detection rate places constraints on the IMF, independent of the assumed simple SFH model. We conclude our likelihood analysis indicates that the data definitively prefers the ``Spock'' galaxy to have a Salpeter IMF ($\alpha = 2.35$) rather than a Top-heavy IMF ($\alpha = 1$) --- which is thought to be the case in the early universe --- with no clear excess of supergiants above the standard IMF.


\end{abstract}



\section{Introduction} \label{sec: intro}

The stellar initial mass function (IMF) describes the number of stars formed in any star-forming episode as a function of stellar mass. Alongside other models/parameters including star-formation history (SFH), metallicity, and dust extinction, the stellar properties of galaxies (such as the 
mass-to-light raio (M/L) and the inferred star-formation rate (SFR)) depend on the IMF  \citep[e.g., ][]{Portinari_2004, McGee_2014, Clauwens_2016}. Moreover, the IMF affects the chemical enrichment process as it determines the relative abundance of massive stars, which release metals into the surroundings via stellar winds during their evolution and when they explode \citep[e.g., ][]{Goswami_2021, Lahen_2024}. The strong ultraviolet (UV) radiation, stellar winds, and supernova explosions associated with massive stars also trigger stellar feedback and affect the subsequent star formation \citep[e.g., ][]{Hennebelle_2008, Hennebelle_2011, Wirth_2022, Dib_2023, Chon_2024, Andersson_2024}. Understanding the IMF at any redshift, for any galaxy, is thus of paramount importance as it provides crucial insights into the formation and evolution of galaxies, as well as the interplay between the cosmic environment and star-formation processes \citep{Bastian2010, Hopkins_2018}.


Canonically, the IMF is characterized as a (broken) power-law distribution, with power-law slope $-\alpha$. In the Milky Way Galaxy, the higher-mass end of the IMF ($M > 1.4\, M_{\odot}$) is found to have $\alpha = 2.35$ through direct star counting of different resolved stellar associations composed of young massive stars \citep[e.g., ][]{Salpeter_1955, MS79, Kroupa_1993}. Such a slope is commonly referred to as the Salpeter IMF. On the other hand, through resolved studies in globular clusters (composed of old stars and therefore confined to lower masses) with different metallicites, the lower-mass end of the IMF ($M < 1.4 M_{\odot}$) is found to have a shallower slope than the Salpeter slope, with variations depending on the correction of binary stars, adopted mass-luminosity relations, and metallicity \citep[e.g., ][]{Kroupa_2001, Chabrier_2003, Li_2023}.

Limited by the spatial resolution of telescopes, measuring the IMF with resolved photometry is only possible for galaxies in the Local Group. Such measurements yield a similar IMF to that measured in the Milky Way \citep[e.g., SMC, LMC, and M31; ][]{Massey_1995, Dario_2009, Lamb_2013, Weisz_2015}. One of the most conventional ways to study the IMF outside of our Local Group (where resolved photometry is unfeasible) is via fitting synthetic spectra to observations \citep{Smith_2020}, where one models how the IMF affects the spectral energy distribution (SED) and/or spectral lines \cite{Vazdekis_2016}. This approach to studying the IMF faces severe degeneracies with other parameters, such as the SFH, metallicity, and dust extinction
\citep[e.g., ][]{Hoversten_2008, Wang_2024}, and furthermore heavily relies on stellar evolution models \citep[e.g., ][]{Ge_2018}. 

Without a robust, direct measurement of what IMF high-redshift galaxies possess, calculations often assume a universal IMF where high-redshift galaxies have the same IMF as local galaxies. A growing body of evidence, however, supports the idea of a variable or evolving IMF \citep[e.g., ][]{Gu_2022, Li_2023}. For example, the recent tension between the masses, brightness, and number density of galaxies at redshifts of $\sim 10$ predicted by $\Lambda$CDM cosmological models and observed by {\it JWST} can be alleviated by a Top-heavy IMF (slope shallower than the locally measured Salpeter one) in the early universe \citep{Haslbauer_2022, 2023NatAs...7..731B,Harikane_2023, woodrum2023jades, Trinca_2024}. On the other hand, diminished dust extinction \citep[e.g., ][]{Ferrara_2023}, bursty star formation \citep[e.g., ][]{sun2023burstystarformationnaturally}, or regulated feedback \citep[e.g., ][]{Dekel_2023} also could resolve the aforementioned tension, removing the need for a Top-heavy IMF at cosmic dawn. Clearly, a more robust diagnostic is needed to measure or constrain the IMF at, especially, the high-mass end.

Here, we examine the feasibility and reliability of a novel way to probe the high-mass end of the IMF at higher redshifts ($z \gtrsim 1$): through gravitational lensing of the most luminous stars belonging to a background galaxy. For this purpose, we use the most massive galaxy clusters known, having masses of $\sim 10^{15} M_{\odot}$, making such clusters the most powerful gravitational lenses. 
In the presence of microlenses comprising intracluster stars, individual stars in galaxies lensed by the galaxy cluster can be further magnified temporarily by an extra factor of hundreds or even thousands \citep{Miralda-Escude_1991, Venumadhav_2017, Diego_2018, Oguri_2018}. Such lensing situations permit individual massive and thus luminous stars at cosmological distances to be temporarily detected under deep imaging from the \textit{Hubble Space Telescope (HST)} and \textit{James Webb Space Telescope (JWST)}. 
The first event of this type is ``Icarus'' ($z = 1.5$), a blue supergiant (BSG) that varies in brightness due to stellar microlensing, as discovered by comparing multiepoch images of the galaxy cluster MACS\,J1149+2223 ($z = 0.54$) \citep{Kelly+18}. Many other similar events were discovered using the same technique employed with {\it HST} \citep[e.g.,][]{rodney18, Chen_2019, Diego2022, Kelly+22, Meena_2023a} and {\it JWST} observations \citep[e.g.,][]{Chen_2022, Meena_2023b, Diego2023_ElGordo, Diego2023_Mothra, Yan+23, fudamoto2024jwst}. The growing number of transient detections in galaxy clusters has made it possible to carry out statistical tests on various astrophysical questions --- for instance, the nature of dark matter --- by probing substructures with the spatial distribution of transients \citep{Dai_2020, Williams+23, Broadhurst2024}.

\begin{figure*}[ht!]
    \centering
    \includegraphics[width = \linewidth]{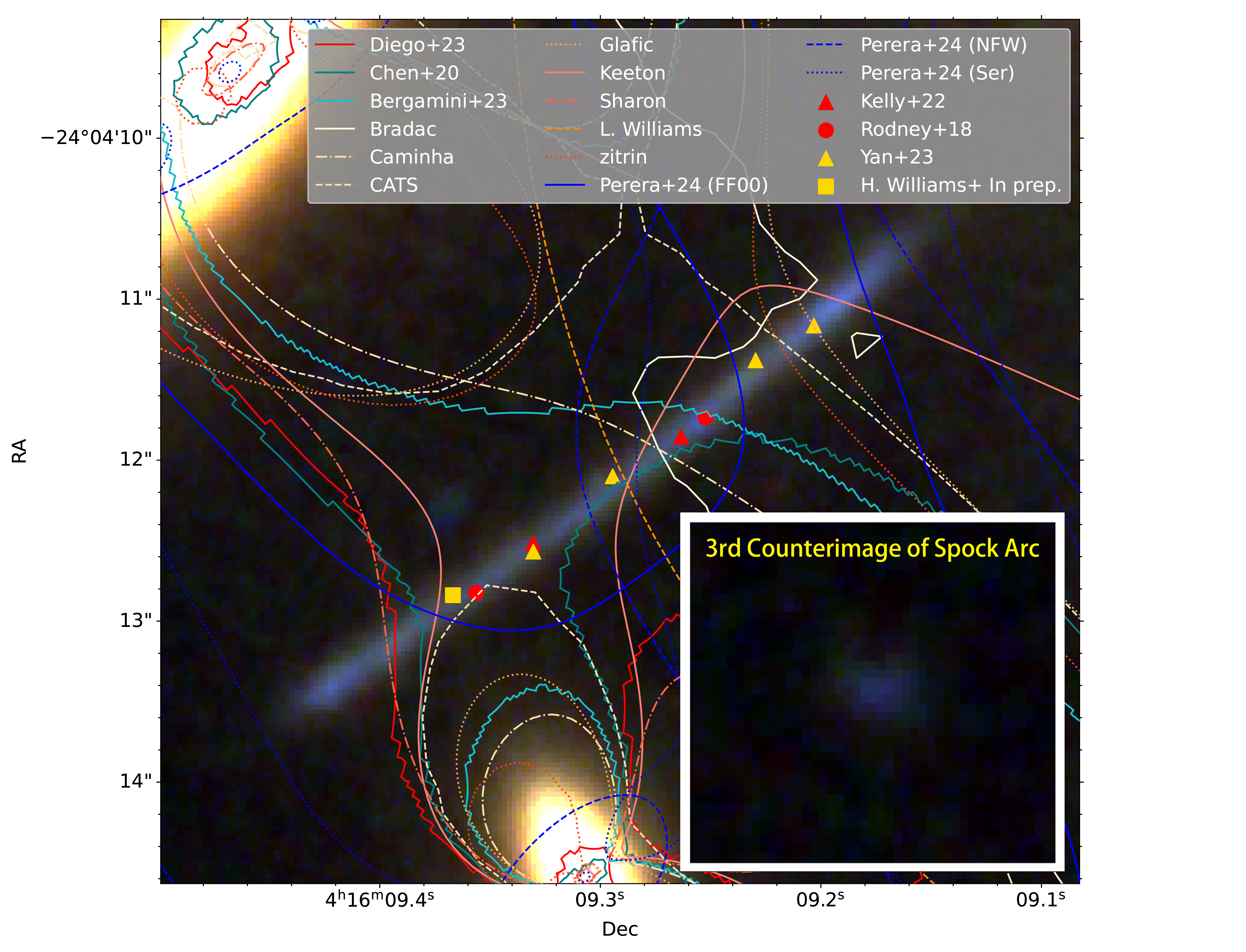}
    \caption{{\it HST} RGB image cutout (Red, F125W + F140W + F160W; Green, F814W + F105W; Blue, F435W + F606W) for the Spock arc. The critical curves (at $z = 1$) of lens models considered in this work are shown in different colors and line styles. The transients reported by \citet{rodney18}, \citet{Kelly+22},  \citet{Yan+23}, and \citet{Williams+inprep} are denoted by different markers, as indicated in the legend. The third counterimage of the ``Spock'' arc is shown in the inset, with the same brightness scaling as the image of the arc. }
    \label{fig: Spock_CC}
\end{figure*}

The detection rate of microlensing transients depends on two major factors: (i) the lensing magnification as imposed by the galaxy cluster as a whole (macrolens) and individual stars in the cluster (microlenses) combined, and (ii) the stellar luminosity function (sLF) of the lensed galaxy. Given that the maximum lensing magnification is limited by the size of the lensed stars \citep[$\sim 10^{2}\,R_\odot$, thus magnification as large as a factor of $\sim 10^{4}$, equivalent to a boost of $\sim 10$ mag; ][]{Miralda-Escude_1991, Oguri_2018}, lensed stars must be intrinsically luminous ($M \lesssim -5$\,mag) to overcome the cosmological distance modulus ($\sim 44$\,mag at $z = 1$) and be detected as a transient. At the shorter wavelength filters (rest-frame UV/optical {\it HST}, and {\it JWST} F090W to F150W), BSGs are the only class of stars that are bright enough to be detected \citep[e.g., ][]{Kelly+18, Chen_2019, Meena_2023a}. They are essentially zero-age main sequence (ZAMS) stars with masses $\gtrsim 20\, M_{\odot}$ and ages of no more than $\sim 5\,$Myr --- thus, their abundance is extremely sensitive to the IMF and the most recent star-formation rate. On the other hand, 
transients detected in the longer wavelength filters (rest-frame IR, {\it JWST} F200W to F444W) are most likely red supergiants (RSGs) \citep[e.g., ][]{diego2023buffaloflashlights}. Unlike BSGs, RSGs are evolved stars in their helium-burning phase, which only lasts $\sim 0.5$--2\,Myr \citep[][]{Ekstr_m_2012, Davies_2017}, with masses in the range $\sim 10$--40\,$M_{\odot}$. Their abundance is thus not directly proportional to the slope of the IMF and is less sensitive to the IMF \citep{Massey_2003, Levesque_2010} compared with BSGs. As we will show in this paper, in light of such a strong correlation between the transient detection rate and the IMF, the transient detection rate can break the IMF-SFH degeneracy and thus place constraints on the IMF when models with different IMFs can give an equally good fit to the observed SED. 






As a case study, we make use of a multiply lensed arc, ``Spock'' ($z = 1.0054$) as shown in Figure~\ref{fig: Spock_CC}, to investigate using the transient detection rate to break the SFH-IMF degeneracy, thus constraining the IMF. We chose this arc as it is one of the arcs with many transient detections in both shorter and longer wavelength filters over the past decade. The galaxy that is being lensed to form the Spock arc (hereafter, the Spock galaxy), is a star-forming galaxy (evident by the strong rest-frame UV flux, shown in Figure~\ref{fig: SED}) lensed by a foreground galaxy cluster, MACS J0416.1-2403 (MACSJ0416, $z = 0.397$). In 2018, two transient events were found in the Spock arc \citep{rodney18} as indicated by red circles in Figure~\ref{fig: Spock_CC}. 
We refer to these two events as the ``Spock'' events in the rest of this paper. On top of the two Spock events, more transient events were discovered on the Spock arc as also labeled in Figure~\ref{fig: Spock_CC}. \citet{Kelly+22} found two transients on the Spock arc by comparing deep two-epoch imaging with the {\it HST} Flashlights survey, as shown by red triangles. 
\citet{Yan+23} found four transients on the Spock Arc in {\it JWST} observations \citep{Willott_2022, Windhorst+23}, as shown by yellow triangles. These four transients were later confirmed by \citet{Williams+inprep}, who found one extra transient from the aforementioned {\it JWST} observations, as denoted by a yellow square. We list all these detections in each of the observed filters in Table~\ref{tab: obs}. 
Throughout this paper, we assume that all these events are in fact stellar microlensing of individual, luminous background stars in the Spock galaxy as none appear to be counterparts of each other, disfavoring the possibility of intrinsic variables \citep{Perera_24}.




This paper is organized as follows. We first introduce the data used in Section~\ref{sec: data}. Our methodology in simulating the transient detection rate in the Spock arc is described in Section~\ref{sec: method}. We present and discuss our results in Sections~\ref{sec: result} and \ref{sec: discussion}, respectively. Finally, we draw our conclusion in Section~\ref{sec: conclusion}. Throughout this paper, we adopt the AB magnitude system \citep{Oke_1983}, along with standard cosmological parameters: $\Omega_{m} = 0.3$, $\Omega_{\Lambda} = 0.7$, and $H_{0} = 70\ \textrm{km s}^{-1}\, \textrm{Mpc}^{-1}$. Under this cosmological model, $1\arcsec = 8.064$\,kpc at $z = 1.0054$, or 5.340\,kpc at $z = 0.396$. The distance modulus at $z = 1.0054$ is 44.11\,mag. The critical surface mass density for a lens of $z = 0.396$ and a source of $z = 1.0054$ would be 2818\,$M_{\odot}$\,pc$^{-2}$. 

\section{Data} \label{sec: data}

\subsection{{\it HST}}

We used image products from two {\it HST} programs.
The first is the Hubble Frontier Field (HFF) program \citep[PI J. Lotz;][]{Lotz_2017}, for which images were taken with the ACS camera in the F435W, F606W, and F814W filters, as well as with the WFC3 camera in the F105W, F125W, F140W, and F160W filters.  We used the final stacked images retrieved from the HFF archive\footnote{\url{https://archive.stsci.edu/prepds/frontier/}}.
The second is the Flashlights Program \citep[PI P. Kelly;][]{Kelly+22}, where images were taken with the WFC, UVIS F200LP, and F350LP, separated by $\sim 1$\,yr. 
The transient detection rate in the {\it HST} filters and their corresponding $5\sigma$ detection limit can be found in Table~\ref{tab: obs}.

\begin{deluxetable}{cccc}
\label{tab: obs}
\centering
\tabletypesize{\scriptsize}
\tablewidth{0pt} 
\tablecaption{Summary of Transients Detected}
\tablehead{Filter  & Number of Transients & Number of & $5\sigma$ Detection\\
& Detected & Pointings & Threshold}
\startdata
\hline
F814W (H) & 2 & 10 & 28.5 \\
F200LP (H)& 2 & 2 & 30\\
F350LP (H)& 2 & 2 & 30\\
F090W (J)& 0 & 4 & 29.7\\
F115W (J)& 0 & 4 & 29.5\\
F150W (J)& 0 & 4 & 29.5\\
F200W (J)& 3 & 4 & 29.5\\
F277W (J)& 5 & 4 & 29.5\\
F356W (J)& 5 & 4 & 29.6\\
F410M (J)& 5 & 4 & 29.0\\
F444W (J)& 4 & 4 & 29.3\\
\enddata
\tablecomments{(1) {\it HST} filters denoted with (H), and {\it JWST} filters denoted with (J)}
\end{deluxetable}

\subsection{{\it JWST}}


We used images from the ``Prime Extragalactic Areas for Reionization and Lensing Science'' (PEARLS; PI R. Windhorst) and the ``CAnadian NIRISS Unbiased Cluster Survey'' (CANUCS; PI C.  Willott) programs observed by {\it JWST}. Four epochs of images were taken (three from PEARLS and one from CANUCS), with NIRCAM filters F090W, F115W, F150W, F200W, F277W, F356W, F410M, and F444W. Observational details are given by \citet{Windhorst+23} and \citet{Willott_2022}, respectively. Based on point-spread-function (PSF) photometry carried out on injected fake sources generated in the latest image calibrations, the corresponding $5\sigma$ depths and the transient detection rate of the aforementioned filters \citep{Williams+inprep} are listed in Table~\ref{tab: obs}.


\section{Methodology} \label{sec: method}


As we have suggested in Section~1, the constraint on the slope of the IMF is placed by reproducing both the SED and transient detection rates together.  To test this, here we adopt a number of different SFH models as well as dust extinction, along with two different IMFs, to fit the SED of the Spock galaxy. We then predict the transient detection rate based on the sLF generated from these SFHs. While almost all these SFHs can fit the SED reasonably well, they predict different transient detection rates depending on the underlying IMF, independent of the underlying SFH model. Therefore, the transient detection rate can be useful in breaking the SFH-IMF degeneracy.

To estimate the transient detection rate, we first define transients as sources that are only detectable in a short period with no signal detected in the same position at other epochs. For simplicity, we refer to individual stars that can potentially be detected as such transients due to microlensing as detectable through microlensing stars \citep[DTM;][]{Diego2024_3M}. This definition distinguishes the transients we focus on from those that are persistent for a long time but vary in brightness 
(e.g., ``Icarus'' studied by \citealt{Kelly+18}, or transients comprising lensed star clusters as studied by \citealt{Dai_2021, Li2024flashlights}). Since DTM stars are undetectable without the magnification boost provided by microlensing, they must be dimmer than the detection threshold when magnified by macromagnification alone. Mathematically, this is represented by the inequality:

\begin{equation}
    m_{f} - 2.5\, \textrm{log}_{10}(\mu_{m}) \ge m_{thr, f}\, ,
\end{equation}

\noindent where $m_{f}$ is the apparent magnitude of a star in filter $f$ if not subject to any lensing magnification, $\mu_{m}$ is the macromagnification brought by the cluster lens, and $m_{thr, f}$ is the $5\sigma$ detection threshold in filter $f$. For the original Spock events, $m_{thr, F814W} = 28.5$\,mag; for events in Flashlights and {\it JWST} pointings, we adopt the $5\sigma$ detection limit introduced earlier in Section~\ref{sec: data}. On the other hand, DTM stars must also be sufficiently bright to be detectable under the effect of microlensing. This is because they could not attain infinitely large magnifications, limited by their sizes \citep{Miralda-Escude_1991, Oguri_2018}. For instance, in a realistic case with a detection threshold of $28.5$\,mag, a macromagnification of $10^{3}$, and a maximum magnification of $10^{4}$ (as boosted by microlensing on top of the macrolens), the corresponding DTM stellar population at $z = 1$ (distance modulus $\sim 44$\,mag) would have absolute magnitudes of $-5.5 \lesssim M \lesssim -8$, corresponding to either BSGs or RSGs. 


Under this definition, we can calculate the expected transient detection rate if we know the abundance of DTM stars in a lensed galaxy (i.e., the sLF), and the probability that these DTM stars can attain sufficient magnification to be temporarily brighter than the detection threshold. In the following subsections, we first describe the way we carry out SED fitting and stellar population synthesis to obtain the sLF (Sec.~\ref{sec: massfunction}). 
We then describe the way to evaluate the probability density function of magnification that a star can attain due to microlensing (hereafter, microlensing PDF) based on lens models (Sec.~\ref{sec: lensing}). We present the calculation of the transient detection rate in Section~\ref{sec: Cal_Rate}. Lastly, we discuss a few important assumptions made in our simulation (Sec.~\ref{sec: Assumptions}).




\subsection{Stellar Luminosity Function}
\label{sec: massfunction}

As emphasized earlier, the inferred transient detection rate is strongly influenced by the sLF. The sLF, in turn, is strongly dependent on various parameters, including age, metallicity, dust, and most importantly, the IMF. The primary way of obtaining these parameters is to carry out SED fitting, where one has to assume specific models and/or parameters. In our calculation, we run different combinations of parameters to fit the SED for two purposes: (1) as a comprehensive examination of whether the reasonable choice of these parameters, in particular the SFH model, is deterministic toward the constraining power of IMF based on transient detection rate; and (2) as a demonstration that SED-fitting alone is degenerate to the SFH-IMF, hence provoking the necessity of using the transient detection rate combined with SED fitting to break the degeneracy.  
We first go through the steps in SED fitting, from retrieving the SED to introducing adopted models/parameters (Sec.~\ref{sec: SED_fitting}); we then describe the way to evolve the SFH derived from SED fitting to obtain the sLF (Sec.~\ref{sec: evolution}).



\begin{deluxetable}{cccc}
\label{tab: Simulation}
\centering
\tabletypesize{\scriptsize}
\tablewidth{0pt} 
\tablecaption{SED Fitting Parameters}
\tablehead{Parameter & Model \textbf{(Index)} & Value/Range}
\startdata
\hline
\hline
Dust    & UV spectral slope \textbf{(1)} & $A_V = 0.6$\\
\citep{Calzetti_00}        & Free Parameter \textbf{(2)} & $0<A_V<1$ \\
        & No Dust \textbf{(3)} & $A_V = 0$ \\
\hline
 & Exponential Decay {\bf (1)}  &  \\
& Total Stellar Mass ($M_{\odot}$) &  $ 10^{4}< M < 10^{9} $\\ 
& Exponential Constant &  $0 < \tau < 1$ \\
& Start Time (Gyr) &  $0 < T_{o} < 2$ \\
& Constant {\bf (2)} &   \\
& Total Stellar Mass ($M_{\odot}$) & $ 10^{4}< M < 10^{9} $ \\ 
SFH model & Start Time (Gyr) &  $0 < T_{o} < 2$ \\
        & Nonparametric {\bf (3)} &   \\
& Stellar Mass in each bin $i$ ($M_{\odot}$) &  $ 10^{4}< M_i < 10^{9} $ \\ 
        & Double Power Law {\bf (4)} &   \\
& Total Stellar Mass ($M_{\odot}$) &  $ 10^{4}< M < 10^{9} $ \\ 
& Falling Slope &  $0 < \alpha < 2$ \\
& Rising Slope &  $0< \beta < 2$ \\ 
& Peak Time (Gyr) &  $0 < \tau < 2$ \\
\hline
IMF & \citet{Salpeter_1955} \textbf{(1)} & $\alpha = 2.35$\\
                     & Top-heavy \textbf{(2)} & $\alpha = 1$\\
\enddata
\tablecomments{We adopt a flat prior for all free parameters.\ 
Numbers in parentheses are the indices used in the ordinate in Figure~\ref{fig: Detection_Rate}.}
\end{deluxetable}

\subsubsection{SED Fitting}
\label{sec: SED_fitting}

\begin{figure}
    \centering
    \includegraphics[width=1\linewidth]{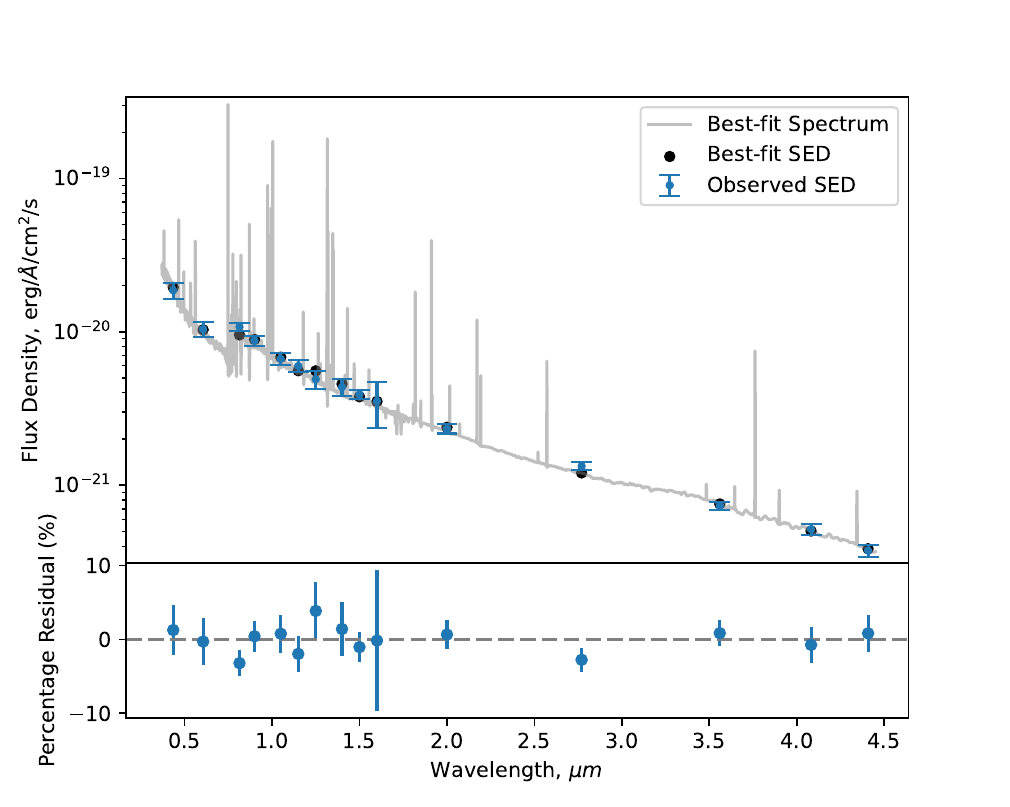}
    \caption{SED of the third counterimage of Spock combining {\it HST} and {\it JWST} images shown as blue data points. For simplicity, we only illustrate the fit for one of the models that adopt a double power law SFH model, a Salpeter IMF with no dust (Index 331) that best fits the observed transient detection rate. We show the best-fit spectrum in gray and the photometry in black data points with residuals in the lower panel. The reduced $\chi^{2}$ of this fit is 0.8, and all the data points can be fitted to within 1--2$\sigma$. }
    \label{fig: SED}
\end{figure}

Considering the lensing geometry, only a small part of the Spock galaxy is being multiply lensed to form the Spock arc. 
Given the similar color (and thus SED) of the third counterimage of the Spock galaxy and the Spock arc itself, here we make use of the third counterimage of the Spock arc (which is far away from the arc itself with no transients detected, as well as isolated from any other potential contaminants, as shown in the inset image in Figure~\ref{fig: Spock_CC}) as in \citet{diego2023buffaloflashlights} to study the stellar population in the Spock arc.
We retrieve its SED from the \citet{Zhang+inprep} catalog \citep[which is constructed via image segmentation with Noise Chisel; ][]{NoiseChisel} and correct it with the lensing magnification ($3.5 \lesssim \mu \lesssim 4$, as predicted by multiple lens models) as shown with blue data points in Figure~\ref{fig: SED}. Figure~\ref{fig: SED} indicates that the SED is extremely bright in the rest-frame UV, demonstrating that there must be a significant young stellar population, some of which must be BSGs.

We carry out SED fitting on the magnification-corrected SED using the SED fitting code {\tt Bagpipes} \citep{carnall18} modified with custom BC03 stellar libraries \citep{BC03}. While we always allow the metallicity and ionization parameters to be freely fitted, SED fitting involves a few critical prior choices of models/parameters that affect the abundance of BSG stars and hence our predicted transient detection rate. They are the SFH model (which describes the SFR, $\Psi (t)$, as a function of time), the dust extinction, and the IMF as summarized in Table~\ref{tab: Simulation}. The SFH models and dust extinction considered are described in detail in Appendix~\ref{sec: append_SFH_detail}. For the IMF, we tested with two distinctive and representative IMFs following the BC03 stellar population library as a simple proof of concept. 
Both IMFs are a single power law: Salpeter IMF ($\alpha = 2.35$) and Top-heavy IMF ($\alpha = 1$). The former is basically a correct representation of the local populations except for the turnover at low masses \citep{Salpeter_1955}, as we reviewed in Section~\ref{sec: intro}. The latter resembles a hypothetical IMF at a higher redshift universe where one expects a shallower IMF with more massive ZAMS stars as motivated by observations \citep[e.g., ][]{Haslbauer_2022, Harikane_2023, Katz_2023, cameron2023nebular}. 
Both IMFs have a mass range of 0.1--100\,$ M_{\odot}$ to be consistent with the mass range allowed by {\tt Bagpipes}. 
Each of the free parameters of the SED fitting is now considered in turn, and we refer the reader to Table~\ref{tab: Sim_details} in the Appendix for detailed information on each of the simulation runs.

\begin{figure}
    \centering
    \includegraphics[width = 1.1\linewidth]{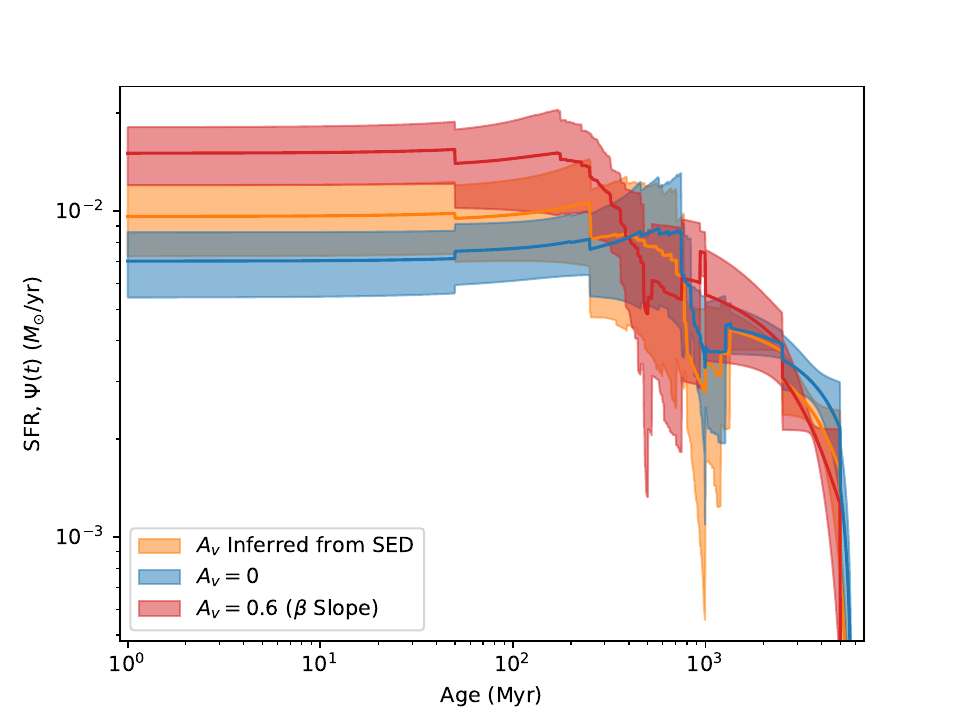}
    \caption{SFH inferred from SED fitting of the third counterimage of the Spock galaxy (shown as an inset in Figure~\ref{fig: Spock_CC}, $\mu = 3.5$), marginalized over the choice of SFH model and IMF. The solid line is the weighted mean of the marginalized SFH, with the band representing $\pm 1\sigma$ of the SFH. The only variable in this figure is the amount of dust, represented by the three colors: (blue) $A_V = 0$; (orange) $A_V = 0.1$--0.3\,mag as obtained from SED fitting; (red) $A_V = 0.6$\,mag. One can see that the choice of SFH model and IMF has a limited effect on the inferred SFH with small scatter --- the dominant factor is the amount of dust, where one can see a clear trend that the inferred SFR is proportional to the amount of dust.}
    \label{fig: all_SFH}
\end{figure}

Iterating through all the combinations of simulation parameters, we obtain $3\, (\textrm{dust}) \times 4\, (\textrm{SFH models})\times 2\, (\textrm{IMFs}) = 24$ sets of SFHs. 
For each of these SFHs, we obtained three parameters from the SED fitting: $\Psi(t)$, $\mathcal{Z}$, and $A_V$. For each combination, we adopt the weighted mean SFH of all the realizations among the nested sampling of {\tt Bagpipes}. The inferred weighted mean SFHs are similar across fits adopting different SFH models/IMFs, and the main difference is caused by the different methods used to estimate dust attenuation. In light of such a convergence, we show the different inferred SFHs in Figure~\ref{fig: all_SFH} marginalized over different choices of SFH models and IMF. Almost all of these combinations of simulation parameters can fit the SED reasonably well. We also note that swapping the IMF between Salpeter/Top-heavy does not affect the fitting result in most cases, consistent with findings in the literature \citep[e.g., ][]{Harvey_2025}. Therefore, it is impossible to distinguish IMFs solely relying on SED fitting, further emphasizing the importance of our methodology in breaking the SFH-IMF degeneracy with the transient detection rate. 

\subsubsection{Stellar Evolution}
\label{sec: evolution}

To obtain the abundance of DTM stars and hence the transient rate for each model, it is necessary to construct the present-day sLF based on the SFH. To do this, we convert the SFHs obtained from SED fitting into multiple small star-formation episodes such that each episode contains a stellar mass of $M_{\star}(t) = \Psi(t) \Delta t$. We adopt a logarithmic step of $\Delta t$ to ensure higher resolution in sampling the young, luminous, but short-lived stars that compose the DTM stellar population (i.e., BSGs and RSGs). We first carry out Monte-Carlo sampling of the ZAMS population in each episode of star formation, following the IMF to be tested. We then evolve each of these episodes based on their age $t$ with a customized {\tt SPISEA} \citep[Stellar Population Interface for Stellar Evolution and Atmospheres][]{SPISEA} following MIST isochrones; \citep[MESA Isochrones \& Stellar Tracks, ][]{Choi_16} while using the metallicity and dust specified for each set of SFHs. To propagate the sampling uncertainty, we repeat the whole process of calculating the sLF 10 times for each SFH. We draw the parameter randomly based on the parameter uncertainties inferred from SED fitting (if applicable; see Table~\ref{tab: Sim_details}) for each run. 

\begin{figure*}
    \centering
    \includegraphics[width=\linewidth]{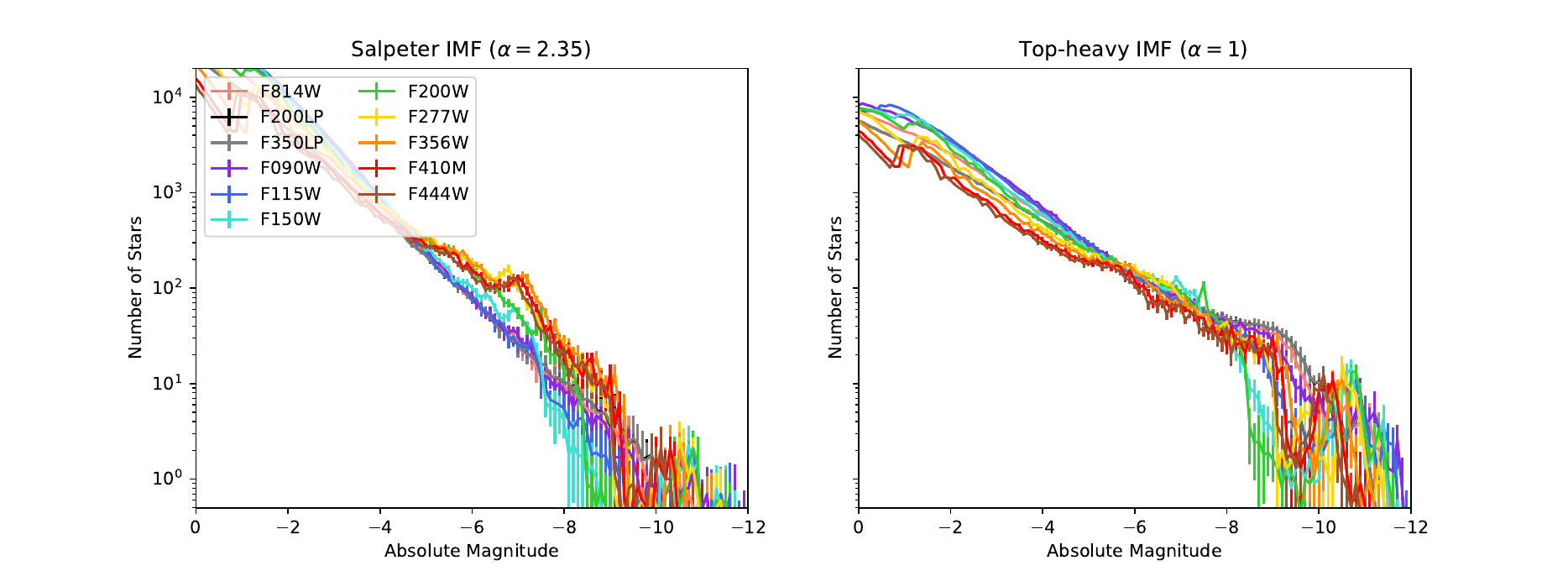}
    \caption{Simulated stellar luminosity function (sLF), marginalized over multiple realizations with error bars representing the standard deviation, based on the third counterimage of the Spock galaxy following the description in Section~\ref{sec: massfunction}, adopting a constant SFH model with no dust, and a Salpeter IMF ($\alpha = 2.35$, left) or a Top-heavy IMF ($\alpha = 1$, right). Different colors represent different filters where transients were detected in the Spock arc, as indicated by the legend. Both SFHs can fit the SED well (reduced $\chi^{2} \approx 1.4$), but one can see that the simulation with a Top-heavy IMF predicts more bright stars than the simulation with a Salpeter IMF at any given filter, as anticipated. This means that they would predict fundamentally different amounts of transients and thus allow one to break the degeneracy. }
    \label{fig: Sim_LF}
\end{figure*}

After evolving all these episodes of star formation, we combine them to obtain the sLF of the Spock galaxy, $N(m_{f})$, for filter $f$. This process is repeated for all 24 sets of SFHs to obtain 24 corresponding sets of sLFs. We show two sets of sLFs generated in Figure~\ref{fig: Sim_LF}; both adopt a constant SFH with no dust, one with a Salpeter IMF (index 321) and one with a Top-heavy IMF (index 322), but provide equally good fits to the SED. It is clearly seen that the simulation with a Top-heavy IMF predicts significantly more bright stars than the simulation with a Salpeter IMF. This means that simulations with a Top-heavy IMF would also give rise to more transients, therefore breaking the degeneracy when both IMFs can fit the SED equally well. Notice that the wiggles towards the end of the sLFs are actually dominated by the isochrone and SFH, as the sampling uncertainty is rather small after marginalizing over the multiple samplings as shown in Figure~\ref{fig: Sim_LF}.


\subsection{Magnification and Microlenses}
\label{sec: lensing}

The next step is to estimate the magnification a star can experience in any random observation. Two major components contribute to the magnification of a single background star: (1) the macromagnification, which is the magnification introduced by the galaxy cluster; and (2) the micromagnification, the magnification produced by a foreground microlens (e.g., intracluster stars). Here, we do not consider the effect of millilenses such as globular clusters embedded in the intracluster light (ICL), which \citet{Diego2024_3M} and \citet{Palencia+24} found to have a limited effect on the overall transient detection rate. We also do not consider the existence of substructures in different proposed forms of dark matter, where it was found to primarily affect the spatial distribution of events instead of the total number of events \citep{Broadhurst2024, Palencia+24}.

\subsubsection{Macrolens}

Different lens models predict different lensing magnifications at different positions as they adopt different constraints and different parameterizations of ingredients, and have different capabilities of reproducing the constraints, as well as different predictive powers. This is particularly true at the position of the Spock arc, where the number of lensing constraints is lower than in other positions of the lens plane. Also, the uncertainty in the mass of nearby member galaxies results in relatively large differences between lens model predictions \citep{Perera_24}. These differences are graphically shown in Figure~\ref{fig: Spock_CC}, where different lens models of MACSJ0416 predict a very different position for the critical curve cutting through the Spock arc. Instead of adopting a single lens model, we carry out our calculation on a total of 14 lens models to mitigate the uncertainty in the predicted magnification in the Spock arc. This includes eight high-resolution lens models available in the HFF archive\footnote{\url{https://archive.stsci.edu/pub/hlsp/frontier/macs0416/models/}} and one recent lens model \citep{Chen_2020} which use multiply lensed images discovered by {\it HST} as constraints, and five recent lens models (\citealt{Diego_MACSJ0416PEARLS}, \citealt{Bergamini_2023}, three in \citealt{Perera_24}) which use multiply lensed images discovered by {\it JWST} as additional constraints.  

Each of the pixels in a lens model represents a certain angular area in the image plane, $\Omega$, and has a specific macromagnification. 
At the scale of a pixel (30\,mas), microlensing would not affect the corresponding source plane area, and the latter only depends on the macromagnification. Since lensing magnification is the ratio between the image plane and the source plane area, we can calculate the source plane area represented by each pixel:

\begin{equation}
\label{eqn: area}
    A_{i} = \frac{\Omega \times (5.34\ \textrm{kpc/\arcsec})^{2}}{|\mu_{m, i}|}\, ,
\end{equation}

\noindent where $A$ and $\mu_{m}$ are the source plane area and macromagnification for pixel $i$, respectively. The value of $\mu_{m}$ can be either $>0$ (outside of the cluster critical curve, known as the positive parity) or $<0$ (inside of the cluster critical curve, known as the negative parity). The statistical behavior of magnification brought by substructure (such as microlenses), as shown in the following section, is different for the two parities \citep{Oguri_2018, Diego_2018, Palencia23, Williams+23, Broadhurst2024}.

\subsubsection{Microlenses}

The contribution of magnification by microlenses is governed by two major components: the macromagnification ($\mu_{m}$) we have just discussed, and the surface mass density of stellar microlenses ($\Sigma_{\star}$). To evaluate the probability of a point source having different magnifications due to microlensing given any macromagnification, we adopt the semianalytical magnification PDF derived by \citet{Palencia23}. These PDFs are extracted from the simulated source-plane magnification (caustic) map combining strong lensing and microlensing effects, and hence capture the probability of a random source of finite size having a certain magnification at any moment. The PDFs only depend on the macromagnification and the abundance of stellar microlenses: 

\begin{equation}
    \label{eqn: palencia_pdf}
    p(\mu; \mu_{t}, \mu_{r}, \Sigma_{\star})\, ,
\end{equation}

\noindent which is a function of the tangential magnification $\mu_{t}$, radial magnification $\mu_{r}$, and $\Sigma_{\star}$. The macromagnification is just the product of the former two parameters, such that for each pixel $i$, we have

\begin{equation}
    \mu_{m, i} = \mu_{t, i} \times \mu_{r}\, .
\end{equation}

For simplicity, we fixed the radial magnification in each model since the Spock arc is a tangential arc where $\mu_{r}$ changes slowly along the arc \citep{diego2023buffaloflashlights}. We adopt $\Sigma_{\star} = 19.45\, M_{\odot}$\,pc$^{-2}$ at the position of the Spock arc as an upper limit of the abundance of intracluster stars as calculated by \citet{rodney18}, who fitted the ICL with an exponential decay SFH model and assuming a \citet{Chabrier_2003} IMF. This value is also close to what one expects from the ICL-dark matter density correlation --- stellar microlenses take up $\sim 2\%$ of the total surface mass density at the position of the Spock arc \citep{Montes_2022, Diego2023_ElGordo,diego2023buffaloflashlights}. 
We shall discuss in Section~\ref{sec: discuss_sigma_star} whether adopting a lower $\Sigma_{\star}$ changes our IMF inference.

\begin{figure*}
    \centering
    \includegraphics[width = 1.1\linewidth]{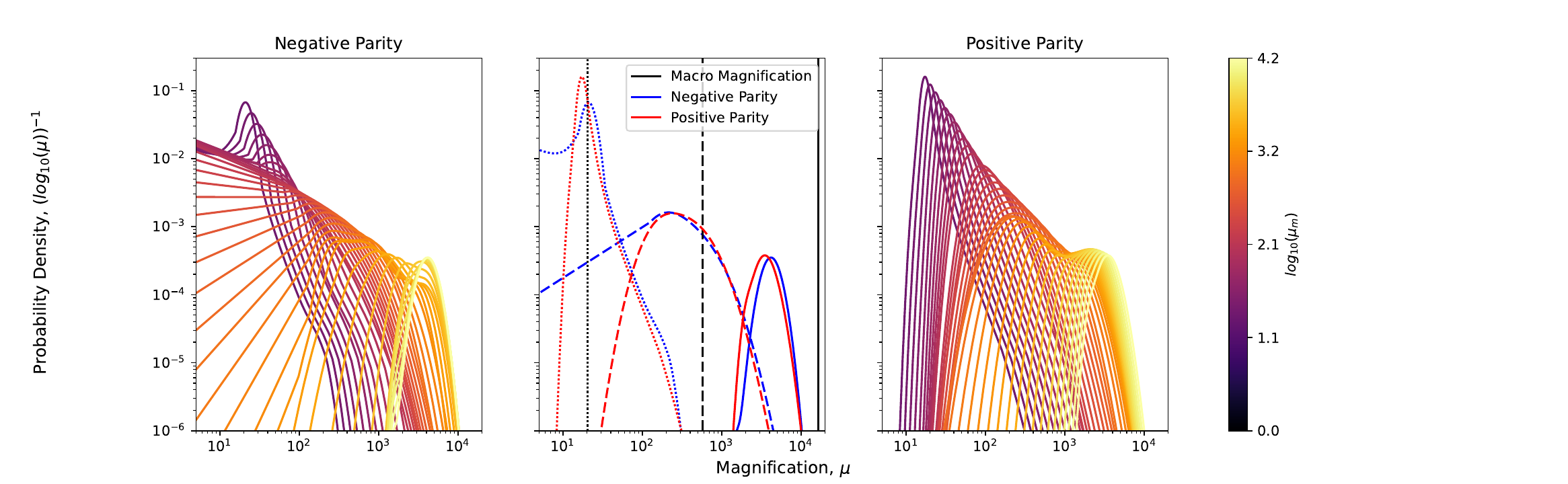}
    \caption{Probability density functions (in logarithmic binning) of magnification induced by the presence of stellar microlenses, given any macromagnification (as shown in different colors), adopting the semianalytical approximation from \citet{Palencia23}. We show the case for negative parity ($\mu_{m} < 0$) on the left and for positive parity ($\mu_{m} > 0$) on the right. As a comparison, we also show the case for a few selected PDFs with the same macromagnification (distinguished by the line style, with the black vertical lines indicating the macromagnification) in the two parities (blue for negative parity;  red for positive parity). For simplicity, we fixed the radial magnification, $\mu_{r}$, to be 1.64 \citep{diego2023buffaloflashlights} and only varied the tangential magnification. We adopt a surface mass density of stellar microlenses of $19.45\,M_{\odot}$\,pc$^{-2}$ --- alternating this value, for example reducing (or increasing) to $10\, M_{\odot}$\,pc$^{-2}$ (or $40\,M_{\odot}$\,pc$^{-2}$), produces a PDF shape similar to taking half (or double) the macromagnification with a slight change in the position of the mode (that depends on the macromagnification). One can also see that the PDFs in both parities converge when $|\mu_{m}|$ is sufficiently large and is extremely optically thick.}
    \label{fig: mu_pdfs}
\end{figure*}

To first order, 
$\Sigma_{\star}$ can also be assumed to be constant along the Spock arc. We thus generate PDFs, $p_{i}(\mu)$, for each pixel in a lens model that is only dependent on the macromagnification ($\mu_{m} \propto \mu_{t}$). As an illustration, we show a selection of PDFs with different macromagnification and different parities in the left and right panels of Figure~\ref{fig: mu_pdfs}, and a small subsample of PDFs with the same macromagnification but different parities in the middle panel. One can see that each of the curves peaks roughly at the adopted macromagnification, except for the case of extremely high macromagnification. Also, those with negative parity have a higher probability of demagnification as well as magnification (both relative to the macromagnification) than the PDFs having the same macromagnification but with positive parity. We return to this issue in Section~\ref{sec: Cal_Rate}.

An immediate caveat of using these PDFs to calculate the transient detection rate is that one would have neglected the correlation between the detection and non-detection of events between shortly separated observations. This is because these PDFs only capture the probability of a background star having any magnification at any random moment, and information regarding the magnification before and after that moment (essentially, the light curve) is lost. If two epochs have short separations, a background star could still be sweeping through the same local caustic network such that the correlation between the (non)detection of the star in consecutive epochs is ignored. A dedicated source plane microlensing simulation is required to resolve such issues, and it is computationally infeasible to do that for our purpose. We describe our measures in minimizing the neglected statistical correlations in Section~\ref{sec: Assumptions}.

\subsection{Predicting Transient Detection Rate}
\label{sec: Cal_Rate}

To calculate the transient detection rate in the Spock arc, given a lens model and an sLF (thus IMF), we evaluate and sum up the transient detection rate in all lens model pixels that represent the arc. As mentioned in the last section, each pixel in a lens model represents a certain source plane area of the background galaxy. Since we carried out SED fitting on the third counterimage, which is the whole Spock galaxy, we assume the distribution of the stars in the Spock galaxy to be homogeneous (such that it does not matter which part of the rather featureless Spock galaxy
gets to be lensed to form the Spock arc).  Given any of the 24 sLFs for the Spock galaxy, $N_{m_{f}}$, the sLF for each pixel is

\begin{equation}
\label{eqn: NumberPerPixel}
    N_{i}(m_{f}) = N(m_{f})\times \frac{A_{i}}{A_{\rm total}}\, ,
\end{equation}

\noindent where $N_{i}(m_{f})$ represents the sLF for pixel $i$ in filter $f$, in terms of apparent magnitude. $A_{\rm total}$ is the source plane area of the whole Spock galaxy, evaluated as $\pi \times (360\, \textrm{pc})^{2}$ as determined by where the light of the third counterimage falls to the background level \citep{diego2023buffaloflashlights}. By adopting Equation~\ref{eqn: NumberPerPixel}, we assume a uniform surface brightness in both the Spock arc and galaxy based on the fact that they are rather featureless, with variation in surface brightness $\lesssim 10\%$ and no significant color differences.

The apparent magnitude of a star after correcting for magnification (i.e., the observed brightness, $m'$), can be written as

\begin{equation}
\label{eqn: MagnifyStar}
    m_{f}' = m_{f} - 2.5\, \textrm{log}_{10}(\mu)\, .
\end{equation}
\noindent
This relation holds for any filter, given that microlensing is monochromatic when the lensed star is a point source, and therefore, the possible spectral variation in the disk of a star can be neglected. In extreme cases, a single star can have a radius up to $\sim 1000\, R_{\odot}$. One part of the star can touch the caustic and attain more magnification than the remaining part of the star, resulting in chromaticity \citep[e.g., ][]{Sajadian_2022}. The larger a star is, the more appreciable this effect becomes. Limited to our use of the PDF, we cannot account for such extreme cases and defer the correction to future work.


With all these ingredients, we can evaluate the expected number of stars in any pixel $i$ that are brighter than the detection threshold at any random moment. This could be done by considering the probability of any star in the area-corrected sLF (Equation~\ref{eqn: NumberPerPixel}) having some specific magnification that is large enough to magnify them to be brighter than the detection limit, given the macromagnification and abundance of microlenses (Equation~\ref{eqn: palencia_pdf}). Mathematically, integrating the number of stars above the detection threshold given the sLF and the microlensing PDF,





\begin{equation}
\label{eqn: raw_rate}
    \mathcal{R}_{i, f} = \int_{-\infty}^{m_{thr, f}} dm_{f} \int_{\mu_{\rm min}}^{\mu_{\rm max}}d\mu  [p_{i}(\mu) N_{i}(m_{f}-2.5\textrm{log}_{10}\mu)]\, , 
\end{equation}

\noindent over the observed brightness and range of magnification ($\mu_{\rm min} = 10^{-1}$ to $\mu_{\rm max} = 10^{4}$) in the PDFs. We truncate at $\mu_{\rm max} = 10^{4}$ because any source with nonzero size has the maximum magnification attained limited by its size, and $10^{4}$ is a characteristic value if we adopt a source size of $\sim 300\, R_{\odot}$ \citep{Palencia23} which is typical for supergiants. The choice of the lower limit follows \citet{Palencia23} as magnification under this limit is not well sampled in their ray-tracing simulation. The statistics lower than this limit are thus not reliable, and they are unimportant because the low magnification never leads to any transient detection.



\begin{figure}
    \centering
    \includegraphics[width = \linewidth]{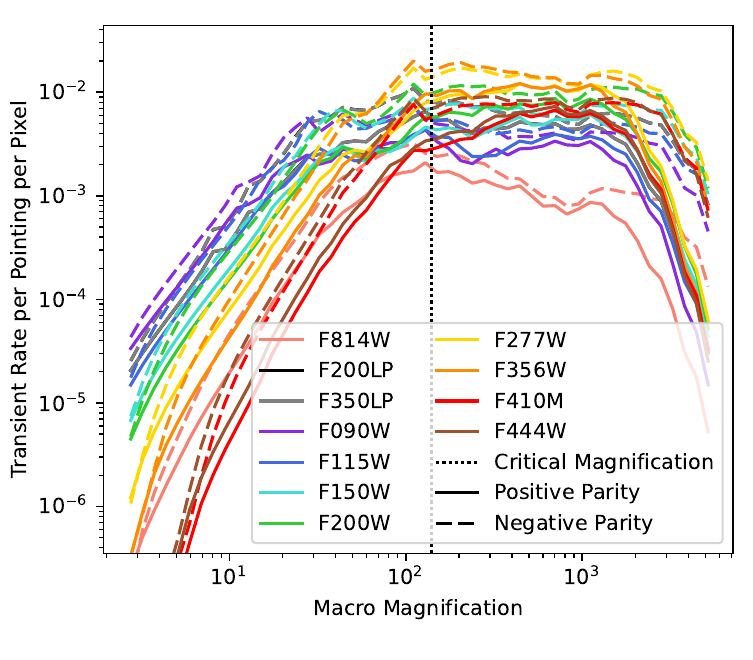}
    \caption{Transient detection rate per pixel per pointing (resolution $0.03\arcsec$) generated with the sLF index {\tt 321} (the one shown in the left panel of Figure~\ref{fig: Sim_LF}), as a function of macromagnification of each pixel. Positive and negative parity are indicated as solid and dashed lines (respectively); the vertical dotted line shows the critical magnification, $\mu_{\textrm{crit}} \approx 140$, where $\mu_{\textrm{crit}} \times \Sigma_{\star} = \Sigma_{\textrm{crit}}$. One can see that the detection rates for most of the filters peak or begin to flatten at the critical magnification, and drop sharply at a magnification of $\sim 2000$ where the reduction of source area wins over the increasing probability of having higher magnification due to microlensing. This is also because the PDFs converge when macromagnification increases, and a further increase in macromagnification no longer increases the probability of having higher magnification.}     
    \label{fig: Sim_MultiFilter}
\end{figure}

Actually, the only difference between any pixels $i$ is their lensing magnification. In other words, we can generate a function $\mathcal{R}_{f}$ that only depends on macromagnification $\mu_{m}$ as shown in Figure~\ref{fig: Sim_MultiFilter}. With $\mathcal{R}_{f}(\mu_{m})$, we can obtain the transient detection rate at pixels with any macromagnification $\mu_{m}$. The exact shape of these functions depends on the chosen sLF, as they have different abundances of DTM stars with different brightness in different filters. However, there are a few common key features of these functions, summarized as follows.
\begin{enumerate}
    \item
The transient detection rate increases quickly with increasing macromagnification in the lower macromagnification regime ($10^{0} \lesssim \mu_{m}  \lesssim 10^{2}$). Although increasing macromagnification decreases the number of stars per pixel owing to the smaller source area probed, the increasing macromagnification quickly increases the probability of having higher magnification with microlensing and thus increases the overall transient detection rate.
    \item 
The increase in transient detection rate slows down and flattens at $\mu_{m} \approx 10^{2}$. At this value of $\mu_{m}$, the microlensing optical depth $\Sigma_{\textrm{eff}} = \mu_{m} \Sigma_{\star} = \Sigma_{\textrm{crit}}$ --- the whole source plane is covered by caustics of microlenses, and the effect of stellar microlensing maximizes. 
At this range of macromagnification, the decrease in the number of bright background stars roughly breaks even with the increase in the probability of having a higher magnification, so the rate flattens. Our simulation aligns with the prediction of \citet{diego2023buffaloflashlights}, where regions with $100 < \mu_{m} < 300$ have the maximum probability of detecting transient stars in F814W with the same abundance of stellar microlenses.
    \item 
The transient detection rate drops quickly in the high macromagnification regime;
$\Sigma_{\textrm{eff}} > \Sigma_{\textrm{crit}}$ and the entire source plane is already covered by microlenses. Adding more microlenses or increasing the magnification no longer increases the size of the demagnification region or the number of microcaustics. Thus, the probability of having higher magnification no longer increases, as illustrated in Figure~\ref{fig: mu_pdfs} where the PDF converges to a log-normal distribution \citep{Palencia23, Diego_2024_Cepheid} and macromagnification increases. This, combined with the decreasing number of background stars per pixel (with increasing high macromagnification), leads to a quick drop in the detection rate. Such an effect is known as ``more-is-less,'' as discussed by \citet{Diego_2019} and \citet{Palencia23}.

    \item 
The detection rate is always higher (factor of $\sim 2$--3) in the negative parity. This is because microlensing gives rise to a higher probability of having larger magnification in the PDFs in the negative parity as shown in Figure~\ref{fig: mu_pdfs}. 
Microlenses always create more demagnification regions in negative parity, and these demagnifications are compensated by having more high-magnification regions such that the total magnification is conserved \citep{Keeton_2003, Oguri_2018, Palencia23}. 
\end{enumerate}

To calculate the transient detection rate per pointing for the entire Spock arc at filter $f$, $\mathcal{R}_{f}$, given a lens model and an sLF (thus depending on the IMF), we can sum up the contributions from all the $\mathcal{N}$ pixels in the lens model (each with a different macromagnification distribution, hence predicting different rates in the end) that represent the arc, as defined by those with signal-to-noise ratio $\ge 5$ in each filter $f$ around the Spock arc:

\begin{equation}
    \mathcal{R}_{f}= \sum_{i = 1}^{\mathcal{N}} \mathcal{R}_{i, f}\, . 
    \label{eqn: resulted_eqn}
\end{equation}



\section{Results} \label{sec: result}

\begin{figure*}
    \centering
    \includegraphics[width = \linewidth]{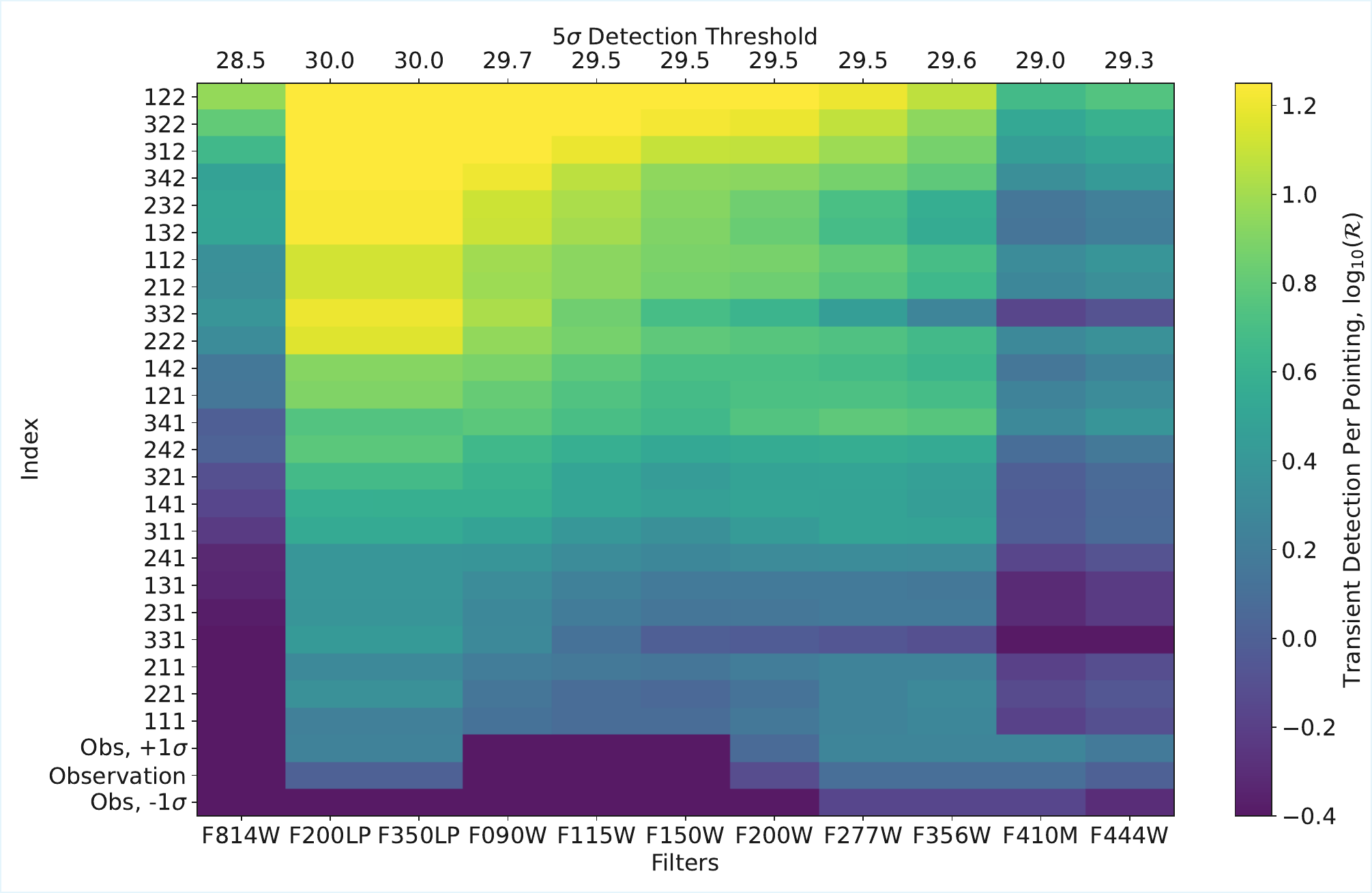}
    \caption{Transient detection rate per pointing ($\mathcal{R}_{f}$, reflected by color in log$_{10}$ scale) for different combinations of simulation parameters (vertical axis, with the first digit representing the amount of dust, the second digit representing the adopted SFH model, and the last digit representing the adopted IMF as given in Table~\ref{tab: Simulation}) in different filters (horizontal axis, with their corresponding $5\sigma$ detection limit in the top axis label) based on the \citet{Perera_24} NFW model. Rows are sorted from the lowest to the highest likelihood compared with the observations (bottom row). As one can see, the best models adopt a Salpeter IMF (last digit ``1''), while the worse models adopt a Top-heavy IMF (last digit ``2''), demonstrating that models adopting a Top-heavy IMF almost always predict significantly more transients than those adopting a Salpeter IMF thus provides a worse fit to the data. The three bottom rows show the observed detection rate and the corresponding $\pm 1\sigma$ range for reference. This allows for a visual comparison that many of the simulations adopting a Salpeter IMF can well reproduce the detection rate in most of the filters; that simulations with a top-heavy IMF always overpredict the detection rate in all the filters.}
    \label{fig: Detection_Rate}
\end{figure*}

With Equation~\ref{eqn: resulted_eqn}, we calculate the expected transient detection rate per pointing in the Spock arc in all filters given an sLF. 
For each lens model, we repeat this calculation for all 24 combinations of simulation parameters listed in Table~\ref{tab: Simulation}. We show the results adopting the \citet[][hereafter P25]{Perera_24} NFW model, which we shall show later best fits the observation, in Figure~\ref{fig: Detection_Rate} as a demonstration. The ordinate of Figure~\ref{fig: Detection_Rate} shows three-digit indices that represent the parameters adopted in each simulation for each row. The three digits are in the order of dust--SFH model--IMF, and follow the numbers in the parentheses of each model shown in Table~\ref{tab: Simulation}. For instance, a simulation with no dust, adopting an exponential decay SFH and a Salpeter IMF, would have an index of {\tt 311}. For reference, we show the observed transient detection rate (and the corresponding $\pm 1\sigma$ Poisson noise) at the bottom of Figure~\ref{fig: Detection_Rate} following Table~\ref{tab: obs}. 

Under the P25 NFW lens model, simulations adopting a Top-heavy IMF (indices ending with ``2'') predict a higher transient detection rate than simulations adopting a Salpeter IMF in all filters. This is expected; stellar populations with a Top-heavy IMF produce more massive DTM stars as shown in Figure~\ref{fig: Sim_LF}, thus predicting a higher transient detection rate. By comparing the predicted rate with the observed rate as shown in Figure~\ref{fig: Detection_Rate}, one can see that under the P25 NFW lens model, most simulations with a Salpeter IMF (indices ending with ``1'') can predict the observation at most filters to a sensible degree (within $\sim 3 \sigma$). In the same bands, simulations with a Top-heavy IMF overpredict the rate by at least an order of magnitude, as also found recently in \citet{Meena_2025}. This immediate result, combined with the fact that almost all SED fittings can produce the SED reasonably well as shown in Section~\ref{sec: massfunction}, demonstrates that the transient detection rate is an effective probe in breaking the SFH-IMF degeneracy, and favors the Spock galaxy having a Salpeter-like IMF rather than a Top-heavy IMF under the P25 NFW lens model.



The aforementioned result is for the P25 NFW lens model --- to take into account the full range of lens models and SFHs, we evaluate the likelihood to quantify which sets of simulations better explain the observations. Since transients are independent events and we always have $\lesssim 1$ event per pointing, they can be considered as Poisson processes ($< 10$ events). We can account for the total detection, $\mathcal{N}$, given $\mathcal{P}$ pointings, such that


\begin{equation}
    \mathcal{N}_{f} = \mathcal{R}_{f} \times \mathcal{P}_{f}\, ,
\end{equation}

\noindent which is guaranteed to be an integer and where $f$ again stands for a given filter. Although the Poissonian function is continuous and one can simply use $\mathcal{R}_{f}$ to calculate the likelihood, we choose to use the total number of transients detected, $\mathcal{N}_{f}$, to weigh the likelihood function toward filters with more pointings (thus having higher statistical significance).
The likelihood function can be written as

\begin{equation}
\label{eqn: likelihood}
    \mathcal{L}_{f} = \frac{(\mathcal{N}_{s, f})^{\mathcal{N}_{o, f}} \times {\rm exp}(-\mathcal{N}_{s, f})}{\mathcal{N}_{o, f}!}\, ,
\end{equation}

\noindent where $\mathcal{N}_{s,f}$ and $\mathcal{N}_{o,f}$ respectively refer to the simulated and observed total number of detections at filter $f$. The joint likelihood of reproducing the observation across all filters, $\mathcal{L}$, is just the product of the likelihoods, $ \prod_{f}^{N} \mathcal{L}_{f}$. 

\begin{deluxetable*}{c|c|c|c|c|c|c|c|c|c|c|c}
\label{tab: list_likelihood}
\centering
\tabletypesize{\scriptsize}
\tablewidth{0pt} 
\tablecaption{Marginalized log$_{10}$ likelihood for lens models, given a Salpeter IMF (Top) or a Top-heavy IMF (Bottom)}
\tablehead{ Lens model & F814W & F200LP & F350LP & F090W & F115W & F150W & F200W & F277W & F356W & F410M & F444W }
\startdata
\citet{Bergamini_2023} & -3.96 & -4.26 & -4.26 & -7.87 & -6.30 & -5.92 & -4.91 & -4.44 & -4.65 & -0.84 & -1.31\\
Bradac & -1.41 & -1.79 & -1.79 & -4.35 & -3.95 & -3.59 & -1.70 & -1.06 & -0.96 & 0.05 & -0.17\\
Caminha & -3.29 & -3.76 & -3.76 & -7.39 & -6.31 & -5.98 & -4.26 & -3.36 & -3.34 & -0.63 & -0.93\\
CATS & -2.68 & -3.12 & -3.12 & -6.02 & -4.96 & -4.59 & -2.99 & -2.39 & -2.22 & -0.44 & -0.70\\
\citet{Chen_2020} & -5.05 & -5.94 & -5.94 & -10.89 & -8.71 & -8.41 & -7.54 & -7.89 & -9.00 & -1.61 & -2.63\\
\citet{Perera_24} (FF00) & -5.67 & -5.87 & -5.86 & -9.68 & -7.55 & -6.96 & -6.88 & -6.64 & -6.98 & -1.19 & -1.97\\
\citet{Perera_24} (NFW) & 0.10* & -0.07* & -0.08* & -1.93* & -1.58* & -1.31* & -0.14* & -0.03* & -0.05* & 0.03 & 0.21*\\
\citet{Perera_24} (Ser) & -0.79 & -1.44 & -1.44 & -4.45 & -4.23 & -3.51 & -1.42 & -0.76 & -0.69 & 0.14 & 0.04\\
\citet{Diego23MACS0416} & -2.83 & -3.64 & -3.64 & -7.37 & -7.15 & -6.71 & -3.83 & -2.15 & -1.61 & -0.19 & -0.44\\
Glafic & -3.33 & -3.52 & -3.52 & -6.56 & -5.20 & -4.83 & -3.85 & -3.33 & -3.26 & -0.63 & -0.92\\
Keeton & -5.71 & -5.87 & -5.87 & -10.79 & -8.44 & -8.68 & -8.92 & -10.91 & -12.89 & -2.60 & -4.06\\
Sharon & -0.34 & -0.64 & -0.64 & -2.52 & -2.85 & -2.72 & -0.75 & -0.26 & -0.17 & -0.02 & 0.20\\
Williams & -0.53 & -1.05 & -1.05 & -3.84 & -3.53 & -2.98 & -1.10 & -0.62 & -0.55 & 0.16* & 0.13\\
Zitrin & -3.90 & -4.13 & -4.13 & -7.5 & -5.81 & -5.37 & -4.55 & -4.12 & -4.27 & -0.78 & -1.20\\
All lens models & 0.35 & 0.08 & 0.08 & -1.82 & -1.55 & -1.28 & 0.01 & 0.30 & 0.34 & 0.89 & 0.86\\
\hline
\hline
\citet{Bergamini_2023} & -18.83 & -16.55 & -16.55 & -27.77 & -25.55 & -27.58 & -24.52 & -16.94 & -11.27 & -3.61 & -5.08\\
Bradac & -8.87 & -9.39 & -9.37 & -17.18 & -14.09 & -13.73 & -11.3 & -7.94 & -4.41 & -0.96 & -1.58\\
Caminha & -16.99 & -15.16 & -15.15 & -25.17 & -23.06 & -24.34 & -23.37 & -15.30 & -9.57 & -2.61 & -3.87\\
CATS & -13.43 & -13.73 & -13.73 & -22.55 & -20.36 & -20.71 & -19.14 & -13.07 & -7.77 & -2.25 & -3.27\\
\citet{Chen_2020} & -25.9 & -20.58 & -20.57 & -35.15 & -34.53 & -39.01 & -30.39 & -21.62 & -15.71 & -5.86 & -7.63\\
\citet{Perera_24} (FF00) & -25.21 & -21.51 & -21.52 & -35.28 & -32.22 & -35.54 & -29.21 & -23.83 & -16.24 & -5.60 & -7.50\\
\citet{Perera_24} (NFW) & -2.75 & -3.33 & -3.32 & -7.82 & -6.52 & -5.90 & -3.42 & -1.64 & -0.90 & 0.06 & -0.13\\
\citet{Perera_24} (Ser) & -7.12 & -8.30 & -8.26 & -15.10 & -13.40 & -12.30 & -8.95 & -5.61 & -2.79 & -0.51 & -0.88\\
\citet{Diego23MACS0416} & -16.45 & -15.78 & -15.74 & -24.61 & -21.17 & -21.05 & -18.58 & -16.32 & -8.95 & -1.45 & -2.47\\
Glafic & -16.42 & -14.46 & -14.45 & -24.02 & -21.06 & -22.7 & -19.99 & -15.51 & -9.80 & -2.68 & -3.98\\
Keeton & -27.68 & -17.95 & -17.95 & -32.18 & -35.21 & -33.01 & -25.94 & -20.57 & -16.16 & -7.00 & -8.70\\
Sharon & -5.17 & -4.89 & -4.88 & -8.90 & -7.58 & -7.72 & -5.10 & -3.26 & -1.88 & -0.06 & -0.18\\
Williams & -5.60 & -7.06 & -7.03 & -13.42 & -11.52 & -10.31 & -7.15 & -4.46 & -2.17 & -0.33 & -0.66\\
Zitrin & -18.4 & -16.28 & -16.28 & -27.29 & -24.41 & -26.23 & -22.32 & -16.74 & -11.17 & -3.52 & -4.96\\
All lens models & -2.75& -3.32 & -3.31 & -7.79 & -6.48 & -5.90 & -3.41 & -1.63 & -0.83 & 0.47 & 0.25\\
\enddata
\tablecomments{The best-fit marginalized lens model for each filter is denoted by *.}
\end{deluxetable*}

To neglect the prior choice of different models/parameters and determine the likelihood for the parameter of interest (in our case, the IMF $\alpha$), one can marginalize the likelihood:

\begin{equation}
\label{eqn: marginalization}
    \mathcal{L}(\alpha)_{\rm marg} = \sum_{\theta} \mathcal{L}(\alpha, \theta)\, .
\end{equation}
\noindent
Here, $\theta$ denotes the parameters over which we marginalize. Since we only care about the effect of the IMF, $\theta$ can include the different amounts of dust, the choice of the SFH model, and lens models. We can also omit one of the $\theta$ from the marginalization to see how the likelihood ratio of IMF preference reacts to the choice of different models, and therefore investigate if the prior choice of some model is deterministic toward the inference. For instance, instead of repeating Figure~\ref{fig: Detection_Rate} for all the lens models, we list the likelihood of all lens models in different filters, marginalized over the choice of SFH models and dust, with the two tested IMFs adopted in Table~\ref{tab: list_likelihood}. The trend observed earlier in Figure~\ref{fig: Detection_Rate} that simulations with Salpeter IMF can better produce the overall detection rate under the P25 NFW lens model, holds for all lens models, as reflected by the likelihood distribution in Table~\ref{tab: list_likelihood}.


With these marginalized likelihoods, we compute the likelihood ratio to determine which IMF is preferred by our simulation:

\begin{equation}
\label{eqn: Bayes_factor}
    K = \frac{\mathcal{L}(\alpha_{\rm SP})_{\textrm{marg}}}{\mathcal{L}(\alpha_{\rm TH})_{\textrm{marg}}} = 10^{47}\, ,
\end{equation}

\noindent which is the ratio of the marginalized likelihood of two competing statistical models (in our case, a Salpeter IMF, $\alpha_{\rm SP}$, versus a Top-heavy IMF, $\alpha_{\rm TH}$). The likelihood ratio of $10^{47}$ means that the likelihood of the model adopting a Salpeter IMF that best reproduces the observations is $\sim 10^{47}$ times larger than the likelihood of the model adopting a Top-heavy IMF that best reproduces the observations. In other words, given our assumptions (including our methodology and parameter space considered), the observations definitively prefer the Spock galaxy to have a Salpeter IMF instead of a Top-heavy IMF. This result aligns with 
\citet{Palencia+24}, 
who find that adopting a Salpeter-like IMF can reasonably reproduce the transient detection rate in the ``Warhol'' arc --- another $z \approx 1$ lensed galaxy in the MACSJ0416 field --- without the need of imposing a shallower IMF.


\section{Discussion} \label{sec: discussion}




In this section, we explore how sensitive our results are to different uncertainties and assumptions. In Section~\ref{sec: effect_SFH}, we inspect whether our choice of SFH model affects the estimation of transient detection rate, and thus if it affects the IMF inference. In Section~\ref{sec: effect_lensmodel}, we examine how the choice of lens model affects our inference for the transient detection rate in the Spock arc. Section~\ref{sec: discuss_sigma_star} discusses how adopting a lower abundance of stellar microlenses affects our IMF inference. Last but not least, we discuss in Section~\ref{sec: Assumptions} various caveats in using microlensing PDFs to calculate the transient detection rate.


\subsection{SFH Model}
\label{sec: effect_SFH}

The simplest quantitative way to evaluate how significantly the choice of the SFH model affects the IMF preference is to exclude the choice of the SFH model from $\theta$ in Equation~\ref{eqn: marginalization}. The corresponding likelihood ratio of which IMF is preferred given an SFH model is then

\begin{equation}
    K(\Psi) = \frac{\mathcal{L}(\alpha_{\rm SP}; \Psi)_{marg}}{\mathcal{L}(\alpha_{\rm TH}; \Psi)_{\rm marg}}\, ,
\end{equation}

\noindent where $\Psi$ denotes the choice of the SFH model. We calculate that $\textrm{log}_{10}(K(\Psi)) = $ 100, 87, 79, and 28 for exponential decay, constant SFH, double power law, and nonparametric SFH model, respectively. Although the likelihood ratios vary across different SFH models, all of the simulations ran with different SFH models definitively prefer the Spock galaxy to have a Salpeter IMF rather than a Top-heavy IMF.

The reason behind this is demonstrated in Figure~\ref{fig: all_SFH} --- the SFH marginalized over different SFH models (as well as IMF) exhibits very small scatter among models with the same amount of dust. Since the SED is dominated by BSGs/RSGs, different SFH models tend to have a very similar $\Psi$ at $t \approx 0$ to fit the SED. The predicted transient detection rate is thus also very similar among the different adopted SFH models. This explains why the choice of SFH is insignificant for the inference of IMF preference.


\subsection{Lens Model}
\label{sec: effect_lensmodel}


\begin{deluxetable}{c|c|c|c}
\label{tab: likelihood}
\centering
\tabletypesize{\scriptsize}
\tablewidth{0pt} 
\tablecaption{Result of Likelihood Analysis}
\tablehead{Lens Model & $\textrm{log}_{10}(\mathcal{L}_{\rm max})$, &  $\textrm{log}_{10}(\mathcal{L}_{\rm max})$,  & likelihood ratio, \\ 
{}& Salpeter& Top-heavy & log$_{10}$($K$)}
\startdata
\citet{Bergamini_2023} & -68.97 & -246.4 & 177.44\\
Bradac & -31.75 & -111.71 & 80.08\\
Caminha & -61.11 & -210.04 & 148.93\\
CATS & -48.82 & -181.68 & 132.86\\
\citet{Chen_2020} & -101.29 & -353.37 & 252.08\\
\citet{Perera_24} (FF00) & -91.64 & -331.73 & 240.09\\
\citet{Perera_24} (NFW) & -13.22 & -39.79 & 26.71\\
\citet{Perera_24} (Ser) & -27.99 & -90.59 & 62.61\\
\citet{Diego23MACS0416} & -56.14 & -174.41 & 118.32\\
Glafic & -56.7 & -202.1 & 145.41\\
Keeton & -115.09 & -384.37 & 269.28\\
Sharon & -18.43 & -52.11 & 33.86\\
Williams & -23.39 & -75.5 & 52.12\\
Zitrin & -65.46 & -237.17 & 171.71\\
\enddata
\tablecomments{(1) Models without specifying the publication year are taken from the HFF Archive. \\ (2) Models prefer Salpeter IMF if log$_{10} K > 0$, Top-heavy IMF if log$_{10} K < 0$}.
\end{deluxetable}

As shown in Figure~\ref{fig: Spock_CC}, lens models predict different numbers of critical curve crossings at different positions, and thus different magnification distributions in the Spock arc. The transient detection rate, which is sensitive to the underlying magnification distribution as demonstrated in Figure~\ref{fig: Sim_MultiFilter}, therefore differs for different lens models. In Table~\ref{tab: list_likelihood}, we showed that it is often different lens models that best reproduce the detection rate in different filters. To investigate the effect of the lens model on inferring IMFs, we computed the maximum joint likelihood assuming a Salpeter or Top-heavy IMF (Equation~\ref{eqn: marginalization}), and the corresponding likelihood ratios (Equation~\ref{eqn: Bayes_factor}) of the simulation adopting different lens models as shown in Table~\ref{tab: likelihood}. 
One can see that the preference for IMF converges among lens models, where all models prefer a Salpeter IMF with the P25 NFW model having the maximum likelihood of $10^{-13.2}$. Nevertheless, our previous marginalization in Equation~\ref{eqn: marginalization} is essentially omitting the choice of lens model, and comparing the likelihood that best reproduces the observation with a Salpeter IMF to that with a Top-heavy IMF i.e., P25 NFW Salpeter model with P25 NFW Top-heavy model). This way, we minimize the effect of the choice of lens models (at least, among those we have considered) and better test if one IMF is preferred over another.
As a side note, there is an extra level of subtlety in this marginalization, because we neglected the fundamental differences between different lens models, for example, they adopt different constraints and have different capabilities in reproducing their constraints as well as predicting different mass distributions \citep{Perera24_MACS0416_Convergence}. Since we only aim to provide a proof of concept in this work, we hereby leave room for refining the marginalization process in future work. 

Another notable feature is that the calculation disfavors lens models with multiple critical curve crossings. From Figure~\ref{fig: Spock_CC}, we can see that three models (two parametric, Keeton and \citealt{Chen_2020}; one nonparametric, P24 FF00) predict multiple crossings in the arc. These three models also have the minimum likelihoods among all the models considered (with either IMF) as shown in Table~\ref{tab: likelihood}, as they over-predict the transient detection rate in general, given an overall larger magnification across the arc. Given that our method can well predict the transient detection rate with some model combinations, this interesting feature perhaps confronts earlier speculations that the Spock arc has multiple critical curve crossings \citep{Diego23MACS0416, Perera_24}.

\subsection{Abundance of Stellar Microlenses}
\label{sec: discuss_sigma_star}

One of the uncertainties is the abundance of stellar microlenses in the position of the Spock arc, as briefly mentioned in Section~\ref{sec: lensing}. Although we have adopted $\Sigma_{\star} = 19.45\, M_{\odot}$\,pc$^{-2}$ throughout the paper, we have also rerun all the calculations adopting $\Sigma_{\star} \approx 5\, M_{\odot}$\,pc$^{-2}$ to test if a smaller amount of stellar microlenses affects our IMF inference. Since there are fewer microlenses, the probability of background stars getting higher magnification is lower and thus decreases the predicted transient detection rate. The immediate consequence is that simulations with a Top-heavy IMF predict fewer transients, albeit still overpredicting, in all the filters. Some of the simulations with the Salpeter IMF now underpredict the detection rate in some filters; however, those that originally overpredicted the number of events now fit the observation well. The likelihood ratio computed for our simulations rerun with $\Sigma_{\star} = 5\, M_{\odot}$\,pc$^{-2}$ is $10^{6.0}$. Even though it is many orders of magnitude lower than the one computed for simulations with $\Sigma_{\star} = 19.45\,M_{\odot}$\,pc$^{-2}$, our simulation still definitively prefers the Salpeter IMF over a Top-heavy IMF with a significantly lower abundance of stellar microlenses.

\subsection{Simulation Assumptions and Limitations}
\label{sec: Assumptions}


By using microlensing PDFs to calculate the transient detection rate, we have made a major assumption that the transient events are not correlated with each other.

One of the possible scenarios where we have omitted the correlation between transient events is that some transients could be counterimages of each other \citep[e.g., Icarus and ``Iapyx'' in][]{Kelly+18}. A single bright star in the background galaxy can be magnified into two (or more) counterimages in the lensed arc, where the two resolved counterimages are being influenced by two different microlenses and thus being detected as two different transients. If treated as two separate background stars, our calculation would overestimate the number of bright stars that can be detected as transients with microlensing, creating a smaller tension with simulations adopting a Top-heavy IMF that overpredicts the detection rate, as we have shown in Section~\ref{sec: result}. That said, \citet{Perera_24} considered the possibility that some of the detected transient events in the Spock arc are counterimages of each other. Assuming different pairs of transients as counterimages of each other and iterating their lens model with the transient ``pairs'' as constraints, they found larger root-mean-square errors in reproducing the lensing constraints. It is hence concluded that none of the transients detected on the Spock arc so far are counterimages of each other, mitigating our concerns in this respect.

Another neglected correlation is the detection and nondetection between simultaneous observations as mentioned in Section~\ref{sec: lensing}. The characteristic caustic-crossing timescale for any microlensing event is $\sim 1$--5 days for a background star with a radius of $\sim 100$--300\,$R_{\odot}$ given a relative transverse velocity of $\sim 500$--1000\,km\,s$^{-1}$ \citep{Kelly+18, Oguri_2018}. If the cadence between two pointings is sufficiently long such that a star would have departed from a single caustic (or a local caustic network for an optically thick case), and the PDF approach remains solid. The time separation between the two Flashlights observations ($\sim 1\,$yr) and the {\it JWST} (PEARLS and CANUCS) pointings ($\ge 14\,$days) are separated by longer than this characteristic timescale, so the PDF method could most likely be safely applied to these detections.  
However, the observational cadence in the SNFrontier program (PI S. Rodney) that led to the discovery of the original Spock events is comparable to this timescale. \citet{diego2023buffaloflashlights} estimated that the Spock events last for $\sim 6\,$days in the observer's frame, over the 30 days of the observation. To minimize the aforementioned caveat of the PDF approach, we follow \citet{diego2023buffaloflashlights} and consider 6\,days as one bin of pointing, deducing a detection rate of $0.2$ transient events per pointing for the Spock events. Nevertheless, this effect would lead to an underestimate of the transient detection rate, providing greater tension with simulations adopting a Top-heavy IMF and not affecting our IMF inference.






\section{Conclusion} \label{sec: conclusion}

The multitude of transients detected in arcs featuring gravitationally lensed high-redshift galaxies revealed the possibility of studying the properties of stellar populations at high redshifts. In this paper, we make use of one lensed galaxy, known as Spock ($z = 1.0054$), with multiple transient detections as shown in Figure~\ref{fig: Spock_CC}, as a proof of concept to demonstrate and examine the possibility of using the transient detection rate as a constraint on the initial mass function (IMF) in distant galaxies. We begin by simulating the stellar luminosity function (sLF, number of stars as a function of their brightness) with simple SED fitting and stellar evolution models (Section~\ref{sec: massfunction}). We then consider the combined effect of strong and microlensing acting (Section~\ref{sec: lensing}) on the sLF to predict the transient detection rate in the Spock arc (Section~\ref{sec: Cal_Rate}). Based on the inference of the transient detection rate considering different simple star formation history (SFH) models and two distinctive IMFs, we achieved the following.



\begin{enumerate}
    \item Investigated how the transient detection rate of a background stellar population depends on the macromagnification provided by strong lens models. As shown in Figure~\ref{fig: Sim_MultiFilter}, we found that the transient detection rate is highest at intermediate macromagnification ($10^{2} \lesssim \mu_{m} \lesssim 10^{3}$). This is where the effect of macromagnification decreasing the number of background stars that could be microlensed is balanced with the increasing probability of background stars having higher magnification with higher macromagnification owing to stellar microlensing.
    
    \item Demonstrated the proof-of-concept methodology and statistical model to probe the IMF of lensed galaxies via the transient detection rate in Section~\ref{sec: method} and \ref{sec: result}. For the case of Spock ($z = 1$), we found that the IMF cannot be distinguished with simple SFH models constrained by SED fitting alone. However, the transient detection rate allows one to break the IMF-SFH degeneracy and thus confront the underlying IMF. A simple likelihood analysis indicates that given our methodology and simple models, the observed transient detection rate definitely prefers the Spock galaxy to have a Salpeter IMF ($\alpha = 2.35$) rather than a Top-heavy IMF ($\alpha = 1$).


    \item 
    Found that the choice of the underlying SFH does not affect the IMF inference, as they all predict a similar most recent star-formation rate (as shown in Figure~\ref{fig: all_SFH}), and therefore a similar transient detection rate.

    \item Found that the choice of lens model is insignificant, as the IMF preference converges among lens models. Nevertheless, our likelihood marginalization allowed us to mitigate the choice of lens model in future cases when one considers less distinctive IMFs, where the effect of the choice of lens model could be more prominent.

\end{enumerate}

We conclude that with an increasing number of transient detections anticipated in the upcoming imaging of galaxy clusters, it is possible to constrain the IMF at earlier
times in the universe
based on our methodology. In this work, we only considered a proof-of-concept bimodal case of Salpeter ($\alpha = 2.35$) versus Top-heavy ($\alpha = 1$) as the IMF of high-redshift lensed galaxies. With more imaging and thus more transients discovered, the increasing signal-to-noise ratio can improve the reliability of the calculation and allow us to evaluate the likelihood distribution in a continuous parameter space of $\alpha$, pinpointing the high-mass end of the IMF to greater precision. 

A possible future direction is to carry out the same practice on higher redshift arcs,
allowing us to place constraints on the IMF as a function of redshift. If we do not find significant evidence of an evolving IMF, then the apparent overabundance of massive galaxies at high redshift might not be solved by an evolving IMF; if we indeed find the IMF evolving, then the timing of the evolution and the way it evolves would provide interesting insights into the cosmic history. Regardless, proceeding in this direction would significantly deepen our understanding of stellar physics in the early universe and potentially revolutionize our comprehension of how the universe evolves. 





\section*{Achknowledgement}

We sincerely thank the anonymous referee for the kind comments that have improved the clarity of this paper.

S.K.L. and J.L. acknowledge support from the Research Grants Council (RGC) of Hong Kong through the General Research Fund (GRF) 17312122. S.K.L., J.L., and T.B.J. are also supported by the Collaborative Research Fund under grant C6017-20G which is issued by the RGC of Hong Kong S.A.R. 
J.M.D. acknowledges support from project PID2022-138896NB-C51 (MCIU/AEI/MINECO/FEDER, UE) Ministerio de Ciencia, Investigaci\'on y Universidades. 
L.W.H.F. acknowledges funding from the UK Science and Technology Facilities Council via grant ST/X001075/1.
J.M.P. received financial support from the Formación de Personal Investigador (FPI) programme, ref. PRE2020-096261, associated with the Spanish Agencia Estatal de Investigaci\'on project MDM-2017-0765-20-2.
A.V.F. is grateful for financial support from the Christopher R. Redlich Fund and numerous donors.
R.A.W. acknowledges support from NASA JWST Interdisciplinary Scientist grants NAG5-12460, NNX14AN10G, and 80NSSC18K0200 from GSFC.

This research is based on observations made with the NASA/ESA {\it Hubble Space Telescope} and {\it James-Webb Space Telescope} obtained from the Space Telescope Science Institute, which is operated by the Association of Universities for Research in Astronomy, Inc., under NASA contract NAS 5–26555. These observations are associated with programs GO-13496, GO-15936, and GO-16278, as well as by GTO-1176 and GTO-1208. 
The data are available at MAST: \dataset[10.17909/gvxn-w422]{\doi{10.17909/gvxn-w422}}.

We acknowledge the usage of the following programs: Astropy \citep{astropy:2013, astropy:2018, astropy:2022}, Numpy \citep{numpy}, Matplotlib \citep{matplotlib} and Scipy \citep{scipy}.

%

\vspace{5mm}
\facilities{{\it HST}, {\it JWST}}


\software{astropy, matplotlib, numpy, scipy}



\appendix

\section{SED Fitting}

Here, we describe in details of the choice of models/parameters as shown in Table~\ref{tab: Simulation}. All the detailed information of each SED fits is shown in Table~\ref{tab: Sim_details}.

\begin{deluxetable*}{c|c|c|c|c|c|c|c|c|c}
\label{tab: Sim_details}
\centering
\tabletypesize{\scriptsize}
\tablewidth{0pt} 
\tablecaption{Complete List of Simulation Parameters}
\tablehead{ID & Index & Dust  & SFH Model & IMF & Metallicity  & log($U$) & Stellar Mass  & Minimum & Number of  \\
& & ($A_V$)& & & ($\mathcal{Z}_{\odot}$) & & ($\textrm{log}_{10} M_{\odot}$) & reduced $\chi^{2}$ & free parameters}
\startdata
1 & 121 & $0.6$ & Constant & Salpeter & $0.4^{+0.2}_{-0.2}$  & $-3.0^{+0.3}_{-0.3}$ & $6.9_{-0.1}^{+0.0}$& 1.4 & 5\\
2 & 122 & $ 0.6$ & Constant & Top Heavy & $0.4^{+0.2}_{-0.2}$  & $-3.0^{+0.3}_{-0.3}$ & $6.9_{-0.1}^{+0.0}$& 1.4 & 5\\
3 & 111 & $ 0.6$ & Exponential Decay & Salpeter & $0.4^{+0.2}_{-0.1}$  & $-3.0^{+0.2}_{-0.4}$ & $6.8_{-0.0}^{+0.1}$& 0.6 & 6\\
4 & 112 & $ 0.6$ & Exponential Decay & Top Heavy & $0.4^{+0.2}_{-0.1}$  & $-3.0^{+0.3}_{-0.4}$ & $6.8_{-0.0}^{+0.1}$& 0.9& 6\\
5 & 141& $ 0.6$ & Non Parametric & Salpeter & $0.3^{+0.1}_{-0.1}$  & $-3.1^{+0.3}_{-0.2}$ & $7.0_{-0.1}^{+0.1}$ & 1.8 & 9\\
6 & 142& $ 0.6$ & Non Parametric & Top Heavy & $0.2^{+0.2}_{-0.1}$  & $-3.1^{+0.3}_{-0.2}$ & $7.0_{-0.1}^{+0.2}$ & 1.8& 9\\
7 & 131& $ 0.6$ & Double Powerlaw & Salpeter & $0.1^{+0.0}_{-0.0}$  & $-3.9^{+0.5}_{-0.4}$ & $7.4_{-0.0}^{+0.0}$ & 5.4&7\\
8 & 132& $ 0.6$ & Double Powerlaw & Top Heavy &  $0.1^{+0.0}_{-0.0}$  & $-3.9^{+0.5}_{-0.4}$ & $7.4_{-0.0}^{+0.0}$ & 5.3&7\\
9 & 221& $  0.3^{+0.1}_{-0.1}$ & Constant & Salpeter &  $0.7^{+0.3}_{-0.4}$ &  $-3.2_{-0.3}^{+0.5}$ & $7.0_{-0.1}^{+0.1}$ & 0.8 & 6\\
10 & 222 & $ 0.3^{+0.1}_{-0.1}$ & Constant & Top Heavy & $0.7^{+0.3}_{-0.4}$  & $-3.2_{-0.3}^{+0.5}$ & $7.0^{+0.0}_{-0.1}$& 0.8& 6\\
11 & 211 & $ 0.1^{+0.2}_{-0.1}$ & Exponential Decay & Salpeter & $0.9^{+0.3}_{-0.3}$  & $-3.3^{+0.4}_{-0.3}$ & $7.0_{-0.1}^{+0.1}$  & 0.8& 7\\
12 & 212 & $ 0.1^{+0.2}_{-0.1}$ & Exponential Decay & Top Heavy & $0.9^{+0.2}_{-0.3}$  & $-3.3^{+0.4}_{-0.3}$ & $7.0_{-0.1}^{+0.1}$ & 0.9& 7\\
13 & 241 & $ 0.1_{-0.1}^{+0.1}$ & Non Parametric & Salpeter & $0.4^{+0.2}_{-0.2}$  & $-3.2^{+0.5}_{-0.4}$ & $7.3_{-0.0}^{+0.1}$ & 1.1& 10\\
14 & 242 & $ 0.1_{-0.1}^{+0.1}$ & Non Parametric & Top Heavy & $0.4^{+0.2}_{-0.2}$  & $-3.2^{+0.5}_{-0.4}$ & $7.3_{-0.1}^{+0.0}$  & 1.1& 10\\
15 & 231 & $ 0.0^{+0.1}_{-0.0}$ & Double Powerlaw & Salpeter &$0.5^{+0.1}_{-0.2}$  & $-3.2^{+0.4}_{-0.4}$ & $7.3_{-0.1}^{+0.0}$ & 1.0&8\\
16 & 232 & $ 0.0^{+0.1}_{-0.0}$ & Double Powerlaw & Top Heavy &  $0.5^{+0.1}_{-0.2}$  & $-3.2^{+0.4}_{-0.4}$ & $7.3_{-0.0}^{+0.1}$ & 1.0&8\\
17 & 321 & $ 0$ & Constant & Salpeter & $1.4^{+0.2}_{-0.2}$  & $-2.9^{+0.2}_{-0.3}$& $7.0^{+0.0}_{-0.0}$ & 1.4 & 5\\
18 & 322 & $ 0$ & Constant & Top Heavy & $1.4^{+0.2}_{-0.2}$  & $-2.9^{+0.2}_{-0.4}$ & $7.0^{+0.0}_{-0.0}$ & 1.4 & 5\\
19 & 311 & $ 0$ & Exponential Decay & Salpeter & $1.1^{+0.3}_{-0.3}$  & $-3.1^{+0.4}_{-0.5}$ & $7.1_{-0.1}^{+0.0}$& 0.9& 6\\
20& 312 & $ 0$ & Exponential Decay & Top Heavy & $1.1^{+0.3}_{-0.3}$  & $-3.1^{+0.4}_{-0.5}$ & $7.1_{-0.1}^{+0.0}$ & 0.9& 6\\
21 & 341 & $ 0$ & Non Parametric & Salpeter & $0.5^{+0.2}_{-0.2}$  & $-3.1^{+0.3}_{-0.4}$ & $7.3_{-0.1}^{+0.1}$ & 1.0& 9\\
22 & 342 & $ 0$ & Non Parametric & Top Heavy & $0.5^{+0.2}_{-0.2}$  & $-3.1^{+0.3}_{-0.5}$ & $7.3_{-0.1}^{+0.1}$& 1.1& 9\\
23 & 331 & $ 0$ & Double Powerlaw & Salpeter & $0.5^{+0.2}_{-0.1}$   & $-3.3^{+0.4}_{-0.3}$ & $7.3_{-0.1}^{+0.0}$ & 0.8&7\\
24 & 332 & $0$ & Double Powerlaw & Top Heavy &  $0.5^{+0.2}_{-0.1}$  & $-3.3^{+0.4}_{-0.3}$ & $7.3_{-0.1}^{+0.0}$ & 0.8&7\\
\enddata
\tablecomments{The fitted parameters are the median, and the upper and lower limits are respectively the 16th and 84th percentiles. }
\end{deluxetable*}

\subsection{SFH models}
\label{sec: append_SFH_detail}


To explore how sensitive our inferences are to different SFHs, we adopted four different simple SFH models available in {\tt Bagpipes} as listed 
in Table~\ref{tab: Simulation}. We have three parametric models --- exponential decay (also known as $\tau$ model), double power law, and constant SFH --- as well as one ``nonparametric'' model where the SFR is allowed to fluctuate with time. For each of the parametric models, we fit simultaneously for (i) $T_{0}$, the time when the star formation begins, such that the star-formation rate, $\Psi(t)$, is always zero for $t < T_{0}$, and (ii) a normalization constant $C$, such that we can solve for the total mass formed, $M$, via a simple integral of  $M = \int^{t}_{0} \Psi(t) dt$.

The simplest model is the constant SFH model,

\begin{equation}
    \Psi(t) = C \, . 
\end{equation}
The exponential decay (or ``$\tau$'') SFH model is 

\begin{equation}
\label{eqn: exp}
    \Psi(t) = C \, \textrm{exp}(- (t-T_{0})/\tau)\, ,
\end{equation}

\noindent where $\tau$ is the decay constant to be freely fitted. The double power law SFH model is expressed as

\begin{equation}
\label{eqn: powerlaw_SFH}
    \Psi(t) = C [(t/\tau)^{\alpha} + (t/\tau)^{-\beta}]^{-1}\, ,
\end{equation}

\noindent where $\tau$ is the time when the SFH peaks, with $\alpha$ and $\beta$ the falling and rising slopes (respectively).

For the nonparametric model, we chose the following time bins similar to \citet{Leja_2019}:
\begin{equation}
    \begin{array}{c}
          \text{$0 \leq t < 10\, \textrm{Myr}$} \\
        \text{$10 \leq t < 50\, \textrm{Myr}$} \\
         \text{$ 50 \leq t < 250\, \textrm{Myr}$} \\
         \text{$250\, \textrm{Myr} \leq t < 1\, \textrm{Gyr}$} \\
          \text{$ 1 \leq t < 2.5\, \textrm{Gyr}$} \\
         \text{$  2.5 \leq t \leq 5\, \textrm{Gyr}$}\, . \\
    \end{array}
\end{equation}
\noindent
The SFR in each time bin is allowed to be a free parameter to be fitted by {\tt Bagpipes}. 

We only adopt SFH models with continuous star formation as opposed to burst models; while the latter is capable of reproducing the SED reasonably well (reduced $\chi^{2} \approx 1$), the resultant models are characterized by ages older than 5\,Myr and hence predict no or very few BSGs 
and underpredict the SW transient detection rate.
Since these SFH models never fit the observed transient detection rate at any SW filters to a sensible degree, we do not include simulations conducted with them in our later analysis, as they would not make a difference.

\subsection{Dust Extinction}

For the dust extinction in the Spock galaxy, we considered three different cases. In the first case, we assume no dust attenuation ($A_V = 0$). The second case is to allow $A_V$ to be one of the free parameters to be solved during SED fitting. Adopting a flat prior of $0 \le A_V \le 1$, the best-fit $A_V$ ranges from 0.1 to 0.3 as shown in Table~\ref{tab: Sim_details}. The last case is to use the UV spectral slope, $\beta$, as an indicator of the dust extinction \citep[][]{Calzetti+94}. We found $\beta_{UV} = -1.54$ from the slope of the SED with the {\it HST} 
F435W and F606W (rest-frame 
$236\,$nm and $288\,$nm) photometry. Assuming $R_V = 3.1$, we convert the UV spectral slope to $A_V = 0.6$\,mag following the correlation found by \citet{reddy18}.

\subsection{Other Free Parameters}

During SED fitting, the metallicity, $\mathcal{Z}$, is freely fitted with a flat prior ($0 \leq Z \leq 3\,Z_{\odot}$). Almost all of our simulations prefer subsolar metallicity, what one would expect for galaxies at $z \approx 1$. For the ionization parameter that describes the degree of ionization in the gas content and thus the strength of emission lines \citep{Carnall_2019}, $\textrm{log}_{10}(U)$, we also allow it to be freely fitted with a flat prior, $-2.5 \geq \textrm{log}_{10}(U) \geq -4.5$, where the nebular metallicity is the same as the stellar metallicity.




\bibliography{sample631}{}

\begin{thebibliography}{}
\expandafter\ifx\csname natexlab\endcsname\relax\def\natexlab#1{#1}\fi
\providecommand{\url}[1]{\href{#1}{#1}}
\providecommand{\dodoi}[1]{doi:~\href{http://doi.org/#1}{\nolinkurl{#1}}}
\providecommand{\doeprint}[1]{\href{http://ascl.net/#1}{\nolinkurl{http://ascl.net/#1}}}
\providecommand{\doarXiv}[1]{\href{https://arxiv.org/abs/#1}{\nolinkurl{https://arxiv.org/abs/#1}}}

\bibitem[{{Akhlaghi} \& {Ichikawa}(2015)}]{NoiseChisel}
{Akhlaghi}, M., \& {Ichikawa}, T. 2015, \apjs, 220, 1, \dodoi{10.1088/0067-0049/220/1/1}

\bibitem[{{Andersson} {et~al.}(2024){Andersson}, {Mac Low}, {Agertz}, {Renaud}, \& {Li}}]{Andersson_2024}
{Andersson}, E.~P., {Mac Low}, M.-M., {Agertz}, O., {Renaud}, F., \& {Li}, H. 2024, \aap, 681, A28, \dodoi{10.1051/0004-6361/202347792}

\bibitem[{{Astropy Collaboration} {et~al.}(2013){Astropy Collaboration}, {Robitaille}, {Tollerud}, {Greenfield}, {Droettboom}, {Bray}, {Aldcroft}, {Davis}, {Ginsburg}, {Price-Whelan}, {Kerzendorf}, {Conley}, {Crighton}, {Barbary}, {Muna}, {Ferguson}, {Grollier}, {Parikh}, {Nair}, {Unther}, {Deil}, {Woillez}, {Conseil}, {Kramer}, {Turner}, {Singer}, {Fox}, {Weaver}, {Zabalza}, {Edwards}, {Azalee Bostroem}, {Burke}, {Casey}, {Crawford}, {Dencheva}, {Ely}, {Jenness}, {Labrie}, {Lim}, {Pierfederici}, {Pontzen}, {Ptak}, {Refsdal}, {Servillat}, \& {Streicher}}]{astropy:2013}
{Astropy Collaboration}, {Robitaille}, T.~P., {Tollerud}, E.~J., {et~al.} 2013, \aap, 558, A33, \dodoi{10.1051/0004-6361/201322068}

\bibitem[{{Astropy Collaboration} {et~al.}(2018){Astropy Collaboration}, {Price-Whelan}, {Sip{\H{o}}cz}, {G{\"u}nther}, {Lim}, {Crawford}, {Conseil}, {Shupe}, {Craig}, {Dencheva}, {Ginsburg}, {Vand erPlas}, {Bradley}, {P{\'e}rez-Su{\'a}rez}, {de Val-Borro}, {Aldcroft}, {Cruz}, {Robitaille}, {Tollerud}, {Ardelean}, {Babej}, {Bach}, {Bachetti}, {Bakanov}, {Bamford}, {Barentsen}, {Barmby}, {Baumbach}, {Berry}, {Biscani}, {Boquien}, {Bostroem}, {Bouma}, {Brammer}, {Bray}, {Breytenbach}, {Buddelmeijer}, {Burke}, {Calderone}, {Cano Rodr{\'\i}guez}, {Cara}, {Cardoso}, {Cheedella}, {Copin}, {Corrales}, {Crichton}, {D'Avella}, {Deil}, {Depagne}, {Dietrich}, {Donath}, {Droettboom}, {Earl}, {Erben}, {Fabbro}, {Ferreira}, {Finethy}, {Fox}, {Garrison}, {Gibbons}, {Goldstein}, {Gommers}, {Greco}, {Greenfield}, {Groener}, {Grollier}, {Hagen}, {Hirst}, {Homeier}, {Horton}, {Hosseinzadeh}, {Hu}, {Hunkeler}, {Ivezi{\'c}}, {Jain}, {Jenness}, {Kanarek}, {Kendrew}, {Kern}, {Kerzendorf}, {Khvalko}, {King}, {Kirkby}, {Kulkarni},
  {Kumar}, {Lee}, {Lenz}, {Littlefair}, {Ma}, {Macleod}, {Mastropietro}, {McCully}, {Montagnac}, {Morris}, {Mueller}, {Mumford}, {Muna}, {Murphy}, {Nelson}, {Nguyen}, {Ninan}, {N{\"o}the}, {Ogaz}, {Oh}, {Parejko}, {Parley}, {Pascual}, {Patil}, {Patil}, {Plunkett}, {Prochaska}, {Rastogi}, {Reddy Janga}, {Sabater}, {Sakurikar}, {Seifert}, {Sherbert}, {Sherwood-Taylor}, {Shih}, {Sick}, {Silbiger}, {Singanamalla}, {Singer}, {Sladen}, {Sooley}, {Sornarajah}, {Streicher}, {Teuben}, {Thomas}, {Tremblay}, {Turner}, {Terr{\'o}n}, {van Kerkwijk}, {de la Vega}, {Watkins}, {Weaver}, {Whitmore}, {Woillez}, {Zabalza}, \& {Astropy Contributors}}]{astropy:2018}
{Astropy Collaboration}, {Price-Whelan}, A.~M., {Sip{\H{o}}cz}, B.~M., {et~al.} 2018, \aj, 156, 123, \dodoi{10.3847/1538-3881/aabc4f}

\bibitem[{{Astropy Collaboration} {et~al.}(2022){Astropy Collaboration}, {Price-Whelan}, {Lim}, {Earl}, {Starkman}, {Bradley}, {Shupe}, {Patil}, {Corrales}, {Brasseur}, {N{\"o}the}, {Donath}, {Tollerud}, {Morris}, {Ginsburg}, {Vaher}, {Weaver}, {Tocknell}, {Jamieson}, {van Kerkwijk}, {Robitaille}, {Merry}, {Bachetti}, {G{\"u}nther}, {Aldcroft}, {Alvarado-Montes}, {Archibald}, {B{\'o}di}, {Bapat}, {Barentsen}, {Baz{\'a}n}, {Biswas}, {Boquien}, {Burke}, {Cara}, {Cara}, {Conroy}, {Conseil}, {Craig}, {Cross}, {Cruz}, {D'Eugenio}, {Dencheva}, {Devillepoix}, {Dietrich}, {Eigenbrot}, {Erben}, {Ferreira}, {Foreman-Mackey}, {Fox}, {Freij}, {Garg}, {Geda}, {Glattly}, {Gondhalekar}, {Gordon}, {Grant}, {Greenfield}, {Groener}, {Guest}, {Gurovich}, {Handberg}, {Hart}, {Hatfield-Dodds}, {Homeier}, {Hosseinzadeh}, {Jenness}, {Jones}, {Joseph}, {Kalmbach}, {Karamehmetoglu}, {Ka{\l}uszy{\'n}ski}, {Kelley}, {Kern}, {Kerzendorf}, {Koch}, {Kulumani}, {Lee}, {Ly}, {Ma}, {MacBride}, {Maljaars}, {Muna}, {Murphy}, {Norman},
  {O'Steen}, {Oman}, {Pacifici}, {Pascual}, {Pascual-Granado}, {Patil}, {Perren}, {Pickering}, {Rastogi}, {Roulston}, {Ryan}, {Rykoff}, {Sabater}, {Sakurikar}, {Salgado}, {Sanghi}, {Saunders}, {Savchenko}, {Schwardt}, {Seifert-Eckert}, {Shih}, {Jain}, {Shukla}, {Sick}, {Simpson}, {Singanamalla}, {Singer}, {Singhal}, {Sinha}, {Sip{\H{o}}cz}, {Spitler}, {Stansby}, {Streicher}, {{\v{S}}umak}, {Swinbank}, {Taranu}, {Tewary}, {Tremblay}, {de Val-Borro}, {Van Kooten}, {Vasovi{\'c}}, {Verma}, {de Miranda Cardoso}, {Williams}, {Wilson}, {Winkel}, {Wood-Vasey}, {Xue}, {Yoachim}, {Zhang}, {Zonca}, \& {Astropy Project Contributors}}]{astropy:2022}
{Astropy Collaboration}, {Price-Whelan}, A.~M., {Lim}, P.~L., {et~al.} 2022, \apj, 935, 167, \dodoi{10.3847/1538-4357/ac7c74}

\bibitem[{{Bastian} {et~al.}(2010){Bastian}, {Covey}, \& {Meyer}}]{Bastian2010}
{Bastian}, N., {Covey}, K.~R., \& {Meyer}, M.~R. 2010, \araa, 48, 339, \dodoi{10.1146/annurev-astro-082708-101642}

\bibitem[{Bergamini {et~al.}(2023)Bergamini, Grillo, Rosati, Vanzella, Meštrić, Mercurio, Acebron, Caminha, Granata, Meneghetti, Angora, \& Nonino}]{Bergamini_2023}
Bergamini, P., Grillo, C., Rosati, P., {et~al.} 2023, Astronomy \&amp; Astrophysics, 674, A79, \dodoi{10.1051/0004-6361/202244834}

\bibitem[{{Boylan-Kolchin}(2023)}]{2023NatAs...7..731B}
{Boylan-Kolchin}, M. 2023, Nature Astronomy, 7, 731, \dodoi{10.1038/s41550-023-01937-7}

\bibitem[{{Broadhurst} {et~al.}(2024){Broadhurst}, {Li}, {Alfred}, {Diego}, {Morilla}, {Kelly}, {Sun}, {Oguri}, {Williams}, {Windhorst}, {Zitrin}, {Abe}, {Chen}, {Fudamoto}, {Kawai}, {Lim}, {Liu}, {Meena}, {Palencia}, {Smoot}, \& {Williams}}]{Broadhurst2024}
{Broadhurst}, T., {Li}, S.~K., {Alfred}, A., {et~al.} 2024, arXiv e-prints, arXiv:2405.19422, \dodoi{10.48550/arXiv.2405.19422}

\bibitem[{{Bruzual} \& {Charlot}(2003)}]{BC03}
{Bruzual}, G., \& {Charlot}, S. 2003, \mnras, 344, 1000, \dodoi{10.1046/j.1365-8711.2003.06897.x}

\bibitem[{{Calzetti} {et~al.}(2000){Calzetti}, {Armus}, {Bohlin}, {Kinney}, {Koornneef}, \& {Storchi-Bergmann}}]{Calzetti_00}
{Calzetti}, D., {Armus}, L., {Bohlin}, R.~C., {et~al.} 2000, \apj, 533, 682, \dodoi{10.1086/308692}

\bibitem[{{Calzetti} {et~al.}(1994){Calzetti}, {Kinney}, \& {Storchi-Bergmann}}]{Calzetti+94}
{Calzetti}, D., {Kinney}, A.~L., \& {Storchi-Bergmann}, T. 1994, \apj, 429, 582, \dodoi{10.1086/174346}

\bibitem[{Cameron {et~al.}(2023)Cameron, Katz, Witten, Saxena, Laporte, \& Bunker}]{cameron2023nebular}
Cameron, A.~J., Katz, H., Witten, C., {et~al.} 2023, Nebular dominated galaxies in the early Universe with top-heavy stellar initial mass functions.
\newblock \doarXiv{2311.02051}

\bibitem[{{Carnall} {et~al.}(2019){Carnall}, {Leja}, {Johnson}, {McLure}, {Dunlop}, \& {Conroy}}]{Carnall_2019}
{Carnall}, A.~C., {Leja}, J., {Johnson}, B.~D., {et~al.} 2019, \apj, 873, 44, \dodoi{10.3847/1538-4357/ab04a2}

\bibitem[{Carnall {et~al.}(2018)Carnall, McLure, Dunlop, \& Dave}]{carnall18}
Carnall, A.~C., McLure, R.~J., Dunlop, J.~S., \& Dave, R. 2018, Monthly Notices of the Royal Astronomical Society, 480, 4379, \dodoi{10.1093/mnras/sty2169}

\bibitem[{{Chabrier}(2003)}]{Chabrier_2003}
{Chabrier}, G. 2003, \pasp, 115, 763, \dodoi{10.1086/376392}

\bibitem[{{Chen} {et~al.}(2020){Chen}, {Kelly}, \& {Williams}}]{Chen_2020}
{Chen}, W., {Kelly}, P.~L., \& {Williams}, L. L.~R. 2020, Research Notes of the American Astronomical Society, 4, 215, \dodoi{10.3847/2515-5172/abcf2b}

\bibitem[{Chen {et~al.}(2019)Chen, Kelly, Diego, Oguri, Williams, Zitrin, Treu, Smith, Broadhurst, Kaiser, Foley, Filippenko, Salo, Hjorth, \& Selsing}]{Chen_2019}
Chen, W., Kelly, P.~L., Diego, J.~M., {et~al.} 2019, The Astrophysical Journal, 881, 8, \dodoi{10.3847/1538-4357/ab297d}

\bibitem[{{Chen} {et~al.}(2022){Chen}, {Kelly}, {Treu}, {Wang}, {Roberts-Borsani}, {Keen}, {Windhorst}, {Zhou}, {Bradac}, {Brammer}, {Strait}, {Broadhurst}, {Diego}, {Frye}, {Meena}, {Zitrin}, {Pascale}, {Castellano}, {Marchesini}, {Morishita}, \& {Yang}}]{Chen_2022}
{Chen}, W., {Kelly}, P.~L., {Treu}, T., {et~al.} 2022, \apjl, 940, L54, \dodoi{10.3847/2041-8213/ac9585}

\bibitem[{{Choi} {et~al.}(2016){Choi}, {Dotter}, {Conroy}, {Cantiello}, {Paxton}, \& {Johnson}}]{Choi_16}
{Choi}, J., {Dotter}, A., {Conroy}, C., {et~al.} 2016, \apj, 823, 102, \dodoi{10.3847/0004-637X/823/2/102}

\bibitem[{{Chon} {et~al.}(2024){Chon}, {Hosokawa}, {Omukai}, \& {Schneider}}]{Chon_2024}
{Chon}, S., {Hosokawa}, T., {Omukai}, K., \& {Schneider}, R. 2024, \mnras, 530, 2453, \dodoi{10.1093/mnras/stae1027}

\bibitem[{{Clauwens} {et~al.}(2016){Clauwens}, {Schaye}, \& {Franx}}]{Clauwens_2016}
{Clauwens}, B., {Schaye}, J., \& {Franx}, M. 2016, \mnras, 462, 2832, \dodoi{10.1093/mnras/stw1808}

\bibitem[{{Da Rio} {et~al.}(2009){Da Rio}, {Gouliermis}, \& {Henning}}]{Dario_2009}
{Da Rio}, N., {Gouliermis}, D.~A., \& {Henning}, T. 2009, \apj, 696, 528, \dodoi{10.1088/0004-637X/696/1/528}

\bibitem[{Dai(2021)}]{Dai_2021}
Dai, L. 2021, Monthly Notices of the Royal Astronomical Society, 501, 5538, \dodoi{10.1093/mnras/stab017}

\bibitem[{Dai \& Miralda-Escudé(2020)}]{Dai_2020}
Dai, L., \& Miralda-Escudé, J. 2020, The Astronomical Journal, 159, 49, \dodoi{10.3847/1538-3881/ab5e83}

\bibitem[{{Davies}(2017)}]{Davies_2017}
{Davies}, B. 2017, Philosophical Transactions of the Royal Society of London Series A, 375, 20160270, \dodoi{10.1098/rsta.2016.0270}

\bibitem[{Dekel {et~al.}(2023)Dekel, Sarkar, Birnboim, Mandelker, \& Li}]{Dekel_2023}
Dekel, A., Sarkar, K.~C., Birnboim, Y., Mandelker, N., \& Li, Z. 2023, Monthly Notices of the Royal Astronomical Society, 523, 3201, \dodoi{10.1093/mnras/stad1557}

\bibitem[{{Dib}(2023)}]{Dib_2023}
{Dib}, S. 2023, \apj, 959, 88, \dodoi{10.3847/1538-4357/ad09bc}

\bibitem[{Diego(2019)}]{Diego_2019}
Diego, J.~M. 2019, Astronomy \& Astrophysics, 625, A84, \dodoi{10.1051/0004-6361/201833670}

\bibitem[{Diego {et~al.}(2022)Diego, {Pascale}, {Kavanagh}, {Kelly}, {Dai}, {Frye}, \& {Broadhurst}}]{Diego2022}
Diego, J.~M., {Pascale}, M., {Kavanagh}, B.~J., {et~al.} 2022, \aap, 665, A134, \dodoi{10.1051/0004-6361/202243605}

\bibitem[{Diego {et~al.}(2024{\natexlab{a}})Diego, {Willner}, {Palencia}, \& {Windhorst}}]{Diego_2024_Cepheid}
Diego, J.~M., {Willner}, S.~P., {Palencia}, J.~M., \& {Windhorst}, R.~A. 2024{\natexlab{a}}, arXiv e-prints, arXiv:2410.09162, \dodoi{10.48550/arXiv.2410.09162}

\bibitem[{Diego {et~al.}(2018)Diego, Kaiser, Broadhurst, Kelly, Rodney, Morishita, Oguri, Ross, Zitrin, Jauzac, Richard, Williams, Vega-Ferrero, Frye, \& Filippenko}]{Diego_2018}
Diego, J.~M., Kaiser, N., Broadhurst, T., {et~al.} 2018, The Astrophysical Journal, 857, 25, \dodoi{10.3847/1538-4357/aab617}

\bibitem[{Diego {et~al.}(2023{\natexlab{a}})Diego, {Meena}, {Adams}, {Broadhurst}, {Dai}, {Coe}, {Frye}, {Kelly}, {Koekemoer}, {Pascale}, {Willner}, {Zackrisson}, {Zitrin}, {Windhorst}, {Cohen}, {Jansen}, {Summers}, {Tompkins}, {Conselice}, {Driver}, {Yan}, {Grogin}, {Marshall}, {Pirzkal}, {Robotham}, {Ryan}, {Willmer}, {Bradley}, {Caminha}, {Caputi}, {Carleton}, \& {Kamieneski}}]{Diego2023_ElGordo}
Diego, J.~M., {Meena}, A.~K., {Adams}, N.~J., {et~al.} 2023{\natexlab{a}}, \aap, 672, A3, \dodoi{10.1051/0004-6361/202245238}

\bibitem[{Diego {et~al.}(2023{\natexlab{b}})Diego, Sun, Yan, Furtak, Zackrisson, Dai, Kelly, Nonino, Adams, Meena, Willner, Zitrin, Cohen, D’Silva, Jansen, Summers, Windhorst, Coe, Conselice, Driver, Frye, Grogin, Koekemoer, Marshall, Pirzkal, Robotham, Rutkowski, Ryan, Tompkins, Willmer, \& Bhatawdekar}]{Diego2023_Mothra}
Diego, J.~M., Sun, B., Yan, H., {et~al.} 2023{\natexlab{b}}, Astronomy \& Astrophysics, 679, A31, \dodoi{10.1051/0004-6361/202347556}

\bibitem[{Diego {et~al.}(2023{\natexlab{c}})Diego, {Adams}, {Willner}, {Harvey}, {Broadhurst}, {Cohen}, {Jansen}, {Summers}, {Windhorst}, {D'Silva}, {Koekemoer}, {Coe}, {Conselice}, {Driver}, {Frye}, {Grogin}, {Marshall}, {Nonino}, {Ortiz}, {Pirzkal}, {Robotham}, {Ryan}, {Willmer}, {Yan}, {Sun}, {Hainline}, {Berkheimer}, {Polletta}, \& {Zitrin}}]{Diego_MACSJ0416PEARLS}
Diego, J.~M., {Adams}, N.~J., {Willner}, S., {et~al.} 2023{\natexlab{c}}, arXiv e-prints, arXiv:2312.11603, \dodoi{10.48550/arXiv.2312.11603}

\bibitem[{Diego {et~al.}(2023{\natexlab{d}})Diego, {Adams}, {Willner}, {Harvey}, {Broadhurst}, {Cohen}, {Jansen}, {Summers}, {Windhorst}, {D'Silva}, {Koekemoer}, {Coe}, {Conselice}, {Driver}, {Frye}, {Grogin}, {Marshall}, {Nonino}, {Ortiz}, {Pirzkal}, {Robotham}, {Ryan}, {Willmer}, {Yan}, {Sun}, {Hainline}, {Berkheimer}, {Polletta}, \& {Zitrin}}]{Diego23MACS0416}
---. 2023{\natexlab{d}}, arXiv e-prints, arXiv:2312.11603, \dodoi{10.48550/arXiv.2312.11603}

\bibitem[{Diego {et~al.}(2024{\natexlab{b}})Diego, {Li}, {Meena}, {Niemiec}, {Acebron}, {Jauzac}, {Struble}, {Amruth}, {Broadhurst}, {Cerny}, {Ebeling}, {Filippenko}, {Jullo}, {Kelly}, {Koekemoer}, {Lagattuta}, {Lim}, {Limousin}, {Mahler}, {Patel}, {Remolina}, {Richard}, {Sharon}, {Steinhardt}, {Umetsu}, {Williams}, {Zitrin}, {Palencia}, {Dai}, {Ji}, \& {Pascale}}]{diego2023buffaloflashlights}
Diego, J.~M., {Li}, S.~K., {Meena}, A.~K., {et~al.} 2024{\natexlab{b}}, \aap, 681, A124, \dodoi{10.1051/0004-6361/202346761}

\bibitem[{Diego {et~al.}(2024{\natexlab{c}})Diego, Li, Amruth, Meena, Broadhurst, Kelly, Filippenko, Williams, Zitrin, Harris, Reina-Campos, Giocoli, Dai, Struble, Treu, Fudamoto, Gilman, Koekemoer, Lim, Palencia, Sun, \& Windhorst}]{Diego2024_3M}
Diego, J.~M., Li, S.~K., Amruth, A., {et~al.} 2024{\natexlab{c}}, Imaging dark matter at the smallest scales with lensed stars.
\newblock \doarXiv{2404.08033}

\bibitem[{Ekström {et~al.}(2012)Ekström, Georgy, Eggenberger, Meynet, Mowlavi, Wyttenbach, Granada, Decressin, Hirschi, Frischknecht, Charbonnel, \& Maeder}]{Ekstr_m_2012}
Ekström, S., Georgy, C., Eggenberger, P., {et~al.} 2012, Astronomy \& Astrophysics, 537, A146, \dodoi{10.1051/0004-6361/201117751}

\bibitem[{Ferrara {et~al.}(2023)Ferrara, Pallottini, \& Dayal}]{Ferrara_2023}
Ferrara, A., Pallottini, A., \& Dayal, P. 2023, Monthly Notices of the Royal Astronomical Society, 522, 3986, \dodoi{10.1093/mnras/stad1095}

\bibitem[{Fudamoto {et~al.}(2024)Fudamoto, Sun, Diego, Dai, Oguri, Zitrin, Zackrisson, Jauzac, Lagattuta, Egami, Iani, Windhorst, Abe, Bauer, Bian, Bhatawdekar, Broadhurst, Cai, Chen, Chen, Cohen, Conselice, Espada, Foo, Frye, Fujimoto, Furtak, Golubchik, Hsiao, Jolly, Kawai, Kelly, Koekemoer, Kohno, Kokorev, Li, Li, Lin, Magdis, Meena, Nabizadeh, Richard, Steinhardt, Wu, Zhu, \& Zou}]{fudamoto2024jwst}
Fudamoto, Y., Sun, F., Diego, J.~M., {et~al.} 2024, JWST Discovery of $40+$ Microlensed Stars in a Magnified Galaxy, the "Dragon" behind Abell 370.
\newblock \doarXiv{2404.08045}

\bibitem[{{Ge} {et~al.}(2018){Ge}, {Yan}, {Cappellari}, {Mao}, {Li}, \& {Lu}}]{Ge_2018}
{Ge}, J., {Yan}, R., {Cappellari}, M., {et~al.} 2018, \mnras, 478, 2633, \dodoi{10.1093/mnras/sty1245}

\bibitem[{{Goswami} {et~al.}(2021){Goswami}, {Slemer}, {Marigo}, {Bressan}, {Silva}, {Spera}, {Boco}, {Grisoni}, {Pantoni}, \& {Lapi}}]{Goswami_2021}
{Goswami}, S., {Slemer}, A., {Marigo}, P., {et~al.} 2021, \aap, 650, A203, \dodoi{10.1051/0004-6361/202039842}

\bibitem[{{Gu} {et~al.}(2022){Gu}, {Greene}, {Newman}, {Kreisch}, {Quenneville}, {Ma}, \& {Blakeslee}}]{Gu_2022}
{Gu}, M., {Greene}, J.~E., {Newman}, A.~B., {et~al.} 2022, \apj, 932, 103, \dodoi{10.3847/1538-4357/ac69ea}

\bibitem[{{Harikane} {et~al.}(2023){Harikane}, {Ouchi}, {Oguri}, {Ono}, {Nakajima}, {Isobe}, {Umeda}, {Mawatari}, \& {Zhang}}]{Harikane_2023}
{Harikane}, Y., {Ouchi}, M., {Oguri}, M., {et~al.} 2023, \apjs, 265, 5, \dodoi{10.3847/1538-4365/acaaa9}

\bibitem[{Harris {et~al.}(2020)Harris, Millman, van~der Walt, Gommers, Virtanen, Cournapeau, Wieser, Taylor, Berg, Smith, Kern, Picus, Hoyer, van Kerkwijk, Brett, Haldane, del R{\'{i}}o, Wiebe, Peterson, G{\'{e}}rard-Marchant, Sheppard, Reddy, Weckesser, Abbasi, Gohlke, \& Oliphant}]{numpy}
Harris, C.~R., Millman, K.~J., van~der Walt, S.~J., {et~al.} 2020, Nature, 585, 357, \dodoi{10.1038/s41586-020-2649-2}

\bibitem[{{Harvey} {et~al.}(2025){Harvey}, {Conselice}, {Adams}, {Austin}, {Juod{\v{z}}balis}, {Trussler}, {Li}, {Ormerod}, {Ferreira}, {Lovell}, {Duan}, {Westcott}, {Harris}, {Bhatawdekar}, {Coe}, {Cohen}, {Caruana}, {Cheng}, {Driver}, {Frye}, {Furtak}, {Grogin}, {Hathi}, {Holwerda}, {Jansen}, {Koekemoer}, {Marshall}, {Nonino}, {Vijayan}, {Wilkins}, {Windhorst}, {Willmer}, {Yan}, \& {Zitrin}}]{Harvey_2025}
{Harvey}, T., {Conselice}, C.~J., {Adams}, N.~J., {et~al.} 2025, \apj, 978, 89, \dodoi{10.3847/1538-4357/ad8c29}

\bibitem[{Haslbauer {et~al.}(2022)Haslbauer, Kroupa, Zonoozi, \& Haghi}]{Haslbauer_2022}
Haslbauer, M., Kroupa, P., Zonoozi, A.~H., \& Haghi, H. 2022, The Astrophysical Journal Letters, 939, L31, \dodoi{10.3847/2041-8213/ac9a50}

\bibitem[{{Hennebelle} \& {Chabrier}(2008)}]{Hennebelle_2008}
{Hennebelle}, P., \& {Chabrier}, G. 2008, \apj, 684, 395, \dodoi{10.1086/589916}

\bibitem[{Hennebelle \& Chabrier(2011)}]{Hennebelle_2011}
Hennebelle, P., \& Chabrier, G. 2011, The Astrophysical Journal Letters, 743, L29, \dodoi{10.1088/2041-8205/743/2/L29}

\bibitem[{{Hopkins}(2018)}]{Hopkins_2018}
{Hopkins}, A.~M. 2018, \pasa, 35, e039, \dodoi{10.1017/pasa.2018.29}

\bibitem[{{Hosek} {et~al.}(2020){Hosek}, {Lu}, {Lam}, {Gautam}, {Lockhart}, {Kim}, \& {Jia}}]{SPISEA}
{Hosek}, Matthew~W., J., {Lu}, J.~R., {Lam}, C.~Y., {et~al.} 2020, \aj, 160, 143, \dodoi{10.3847/1538-3881/aba533}

\bibitem[{{Hoversten} \& {Glazebrook}(2008)}]{Hoversten_2008}
{Hoversten}, E.~A., \& {Glazebrook}, K. 2008, \apj, 675, 163, \dodoi{10.1086/524095}

\bibitem[{Hunter(2007)}]{matplotlib}
Hunter, J.~D. 2007, Computing in Science \& Engineering, 9, 90, \dodoi{10.1109/MCSE.2007.55}

\bibitem[{Katz {et~al.}(2023)Katz, Kimm, Ellis, Devriendt, \& Slyz}]{Katz_2023}
Katz, H., Kimm, T., Ellis, R.~S., Devriendt, J., \& Slyz, A. 2023, Monthly Notices of the Royal Astronomical Society, 524, 351–360, \dodoi{10.1093/mnras/stad1903}

\bibitem[{Keeton(2003)}]{Keeton_2003}
Keeton, C.~R. 2003, The Astrophysical Journal, 584, 664, \dodoi{10.1086/345717}

\bibitem[{Kelly {et~al.}(2018)Kelly, Diego, Rodney, Kaiser, Broadhurst, Zitrin, Treu, Perez-Gonzalez, Morishita, Jauzac, Selsing, Oguri, Pueyo, Ross, Filippenko, Smith, Hjorth, Cenko, Wang, Howell, Richard, Frye, Jha, Foley, Norman, Bradac, Zheng, Brammer, Benito, Cava, Christensen, de~Mink, Graur, Grillo, Kawamata, Kneib, Matheson, McCully, Nonino, Perez-Fournon, Riess, Rosati, Schmidt, Sharon, \& Weiner}]{Kelly+18}
Kelly, P.~L., Diego, J.~M., Rodney, S., {et~al.} 2018, Extreme magnification of a star at redshift 1.5 by a galaxy-cluster lens.
\newblock \doarXiv{1706.10279}

\bibitem[{{Kelly} {et~al.}(2022){Kelly}, {Chen}, {Alfred}, {Broadhurst}, {Diego}, {Emami}, {Filippenko}, {Keen}, {Li}, {Lim}, {Meena}, {Oguri}, {Scarlata}, {Treu}, {Williams}, {Williams}, {Zhou}, {Zitrin}, {Foley}, {Jha}, {Kaiser}, {Mehta}, {Rieck}, {Salo}, {Smith}, \& {Weisz}}]{Kelly+22}
{Kelly}, P.~L., {Chen}, W., {Alfred}, A., {et~al.} 2022, arXiv e-prints, arXiv:2211.02670, \dodoi{10.48550/arXiv.2211.02670}

\bibitem[{Kroupa(2001)}]{Kroupa_2001}
Kroupa, P. 2001, Monthly Notices of the Royal Astronomical Society, 322, 231, \dodoi{10.1046/j.1365-8711.2001.04022.x}

\bibitem[{{Kroupa} {et~al.}(1993){Kroupa}, {Tout}, \& {Gilmore}}]{Kroupa_1993}
{Kroupa}, P., {Tout}, C.~A., \& {Gilmore}, G. 1993, \mnras, 262, 545, \dodoi{10.1093/mnras/262.3.545}

\bibitem[{{Lah{\'e}n} {et~al.}(2024){Lah{\'e}n}, {Naab}, \& {Sz{\'e}csi}}]{Lahen_2024}
{Lah{\'e}n}, N., {Naab}, T., \& {Sz{\'e}csi}, D. 2024, \mnras, 530, 645, \dodoi{10.1093/mnras/stae904}

\bibitem[{{Lamb} {et~al.}(2013){Lamb}, {Oey}, {Graus}, {Adams}, \& {Segura-Cox}}]{Lamb_2013}
{Lamb}, J.~B., {Oey}, M.~S., {Graus}, A.~S., {Adams}, F.~C., \& {Segura-Cox}, D.~M. 2013, \apj, 763, 101, \dodoi{10.1088/0004-637X/763/2/101}

\bibitem[{{Leja} {et~al.}(2019){Leja}, {Carnall}, {Johnson}, {Conroy}, \& {Speagle}}]{Leja_2019}
{Leja}, J., {Carnall}, A.~C., {Johnson}, B.~D., {Conroy}, C., \& {Speagle}, J.~S. 2019, \apj, 876, 3, \dodoi{10.3847/1538-4357/ab133c}

\bibitem[{{Levesque}(2010)}]{Levesque_2010}
{Levesque}, E.~M. 2010, in Astronomical Society of the Pacific Conference Series, Vol. 425, Hot and Cool: Bridging Gaps in Massive Star Evolution, ed. C.~{Leitherer}, P.~D. {Bennett}, P.~W. {Morris}, \& J.~T. {Van Loon}, 103, \dodoi{10.48550/arXiv.0911.4720}

\bibitem[{Li {et~al.}(2023)Li, Liu, Zhang, Tian, Fu, Li, \& Yan}]{Li_2023}
Li, J., Liu, C., Zhang, Z.-Y., {et~al.} 2023, Nature, 613, 460–462, \dodoi{10.1038/s41586-022-05488-1}

\bibitem[{Li {et~al.}(2024)Li, Kelly, Diego, Lim, Chen, Alfred, Williams, Broadhurst, Meena, Zitrin, \& Chow}]{Li2024flashlights}
Li, S.~K., Kelly, P.~L., Diego, J.~M., {et~al.} 2024, Flashlights: Microlensing vs Stellar Variability of Transients in the Star Clusters of the Dragon Arc.
\newblock \doarXiv{2404.08571}

\bibitem[{{Lotz} {et~al.}(2017){Lotz}, {Koekemoer}, {Coe}, {Grogin}, {Capak}, {Mack}, {Anderson}, {Avila}, {Barker}, {Borncamp}, {Brammer}, {Durbin}, {Gunning}, {Hilbert}, {Jenkner}, {Khandrika}, {Levay}, {Lucas}, {MacKenty}, {Ogaz}, {Porterfield}, {Reid}, {Robberto}, {Royle}, {Smith}, {Storrie-Lombardi}, {Sunnquist}, {Surace}, {Taylor}, {Williams}, {Bullock}, {Dickinson}, {Finkelstein}, {Natarajan}, {Richard}, {Robertson}, {Tumlinson}, {Zitrin}, {Flanagan}, {Sembach}, {Soifer}, \& {Mountain}}]{Lotz_2017}
{Lotz}, J.~M., {Koekemoer}, A., {Coe}, D., {et~al.} 2017, \apj, 837, 97, \dodoi{10.3847/1538-4357/837/1/97}

\bibitem[{{Massey} {et~al.}(1995){Massey}, {Lang}, {Degioia-Eastwood}, \& {Garmany}}]{Massey_1995}
{Massey}, P., {Lang}, C.~C., {Degioia-Eastwood}, K., \& {Garmany}, C.~D. 1995, \apj, 438, 188, \dodoi{10.1086/175064}

\bibitem[{{Massey} \& {Olsen}(2003)}]{Massey_2003}
{Massey}, P., \& {Olsen}, K.~A.~G. 2003, \aj, 126, 2867, \dodoi{10.1086/379558}

\bibitem[{McGee {et~al.}(2014)McGee, Goto, \& Balogh}]{McGee_2014}
McGee, S.~L., Goto, R., \& Balogh, M.~L. 2014, Monthly Notices of the Royal Astronomical Society, 438, 3188–3204, \dodoi{10.1093/mnras/stt2426}

\bibitem[{Meena {et~al.}(2023{\natexlab{a}})Meena, Chen, Zitrin, Kelly, Golubchik, Zhou, Alfred, Broadhurst, Diego, Filippenko, Li, Oguri, Smith, \& Williams}]{Meena_2023a}
Meena, A.~K., Chen, W., Zitrin, A., {et~al.} 2023{\natexlab{a}}, Monthly Notices of the Royal Astronomical Society, 521, 5224, \dodoi{10.1093/mnras/stad869}

\bibitem[{Meena {et~al.}(2023{\natexlab{b}})Meena, Zitrin, Jiménez-Teja, Zackrisson, Chen, Coe, Diego, Dimauro, Furtak, Kelly, Oguri, Welch, Andrade-Santos, Adamo, Bhatawdekar, Bradač, Bradley, Broadhurst, Conselice, Dayal, Donahue, Frye, Fujimoto, Hsiao, Kokorev, Mahler, Vanzella, \& Windhorst}]{Meena_2023b}
Meena, A.~K., Zitrin, A., Jiménez-Teja, Y., {et~al.} 2023{\natexlab{b}}, The Astrophysical Journal Letters, 944, L6, \dodoi{10.3847/2041-8213/acb645}

\bibitem[{{Meena} {et~al.}(2025){Meena}, {Li}, {Zitrin}, {Kelly}, {Broadhurst}, {Chen}, {Diego}, {Filippenko}, {Furtak}, \& {Williams}}]{Meena_2025}
{Meena}, A.~K., {Li}, S.~K., {Zitrin}, A., {et~al.} 2025, arXiv e-prints, arXiv:2503.21706, \dodoi{10.48550/arXiv.2503.21706}

\bibitem[{{Miller} \& {Scalo}(1979)}]{MS79}
{Miller}, G.~E., \& {Scalo}, J.~M. 1979, \apjs, 41, 513, \dodoi{10.1086/190629}

\bibitem[{{Miralda-Escude}(1991)}]{Miralda-Escude_1991}
{Miralda-Escude}, J. 1991, \apj, 379, 94, \dodoi{10.1086/170486}

\bibitem[{{Montes}(2022)}]{Montes_2022}
{Montes}, M. 2022, Nature Astronomy, 6, 308, \dodoi{10.1038/s41550-022-01616-z}

\bibitem[{Oguri {et~al.}(2018)Oguri, Diego, Kaiser, Kelly, \& Broadhurst}]{Oguri_2018}
Oguri, M., Diego, J.~M., Kaiser, N., Kelly, P.~L., \& Broadhurst, T. 2018, Phys. Rev. D, 97, 023518, \dodoi{10.1103/PhysRevD.97.023518}

\bibitem[{{Oke} \& {Gunn}(1983)}]{Oke_1983}
{Oke}, J.~B., \& {Gunn}, J.~E. 1983, \apj, 266, 713, \dodoi{10.1086/160817}

\bibitem[{Palencia {et~al.}(2024)Palencia, Diego, \& Kavanagh}]{Palencia23}
Palencia, J.~M., Diego, J.~M., \& Kavanagh, B. J.and Martínez-Arrizabalaga, J. 2024, A\&A, 687, A81, \dodoi{10.1051/0004-6361/202347492}

\bibitem[{Palencia {et~al.}(2025)Palencia, Diego, Dai, Pascale, Windhorst, Koekemoer, Li, Kavanagh, Sun, Alfred, Meena, Broadhurst, Kelly, Perera, Williams, \& Zitrin}]{Palencia+24}
Palencia, J.~M., Diego, J.~M., Dai, L., {et~al.} 2025, Microlensing at Cosmological Distances: Event Rate Predictions in the Warhol Arc of MACS 0416.
\newblock \doarXiv{2504.07039}

\bibitem[{{Perera} {et~al.}(2024){Perera}, {Miller}, {Williams}, {Liesenborgs}, {Keen}, {Li}, \& {Limousin}}]{Perera24_MACS0416_Convergence}
{Perera}, D., {Miller}, John~H, J., {Williams}, L. L.~R., {et~al.} 2024, arXiv e-prints, arXiv:2411.05083, \dodoi{10.48550/arXiv.2411.05083}

\bibitem[{{Perera} {et~al.}(2025){Perera}, {Williams}, {Liesenborgs}, {Kelly}, {Taft}, {Li}, {Jauzac}, {Diego}, {Natarajan}, {Steinhardt}, {Faisst}, {Rich}, \& {Limousin}}]{Perera_24}
{Perera}, D., {Williams}, L. L.~R., {Liesenborgs}, J., {et~al.} 2025, \mnras, 536, 2690, \dodoi{10.1093/mnras/stae2753}

\bibitem[{{Portinari} {et~al.}(2004){Portinari}, {Sommer-Larsen}, \& {Tantalo}}]{Portinari_2004}
{Portinari}, L., {Sommer-Larsen}, J., \& {Tantalo}, R. 2004, \mnras, 347, 691, \dodoi{10.1111/j.1365-2966.2004.07207.x}

\bibitem[{{Reddy} {et~al.}(2018){Reddy}, {Shapley}, {Sanders}, {Kriek}, {Coil}, {Shivaei}, {Freeman}, {Mobasher}, {Siana}, {Azadi}, {Fetherolf}, {Fornasini}, {Leung}, {Price}, {Zick}, \& {Barro}}]{reddy18}
{Reddy}, N.~A., {Shapley}, A.~E., {Sanders}, R.~L., {et~al.} 2018, \apj, 869, 92, \dodoi{10.3847/1538-4357/aaed1e}

\bibitem[{{Rodney} {et~al.}(2018){Rodney}, {Balestra}, {Bradac}, {Brammer}, {Broadhurst}, {Caminha}, {Chiriv{\i}}, {Diego}, {Filippenko}, {Foley}, {Graur}, {Grillo}, {Hemmati}, {Hjorth}, {Hoag}, {Jauzac}, {Jha}, {Kawamata}, {Kelly}, {McCully}, {Mobasher}, {Molino}, {Oguri}, {Richard}, {Riess}, {Rosati}, {Schmidt}, {Selsing}, {Sharon}, {Strolger}, {Suyu}, {Treu}, {Weiner}, {Williams}, \& {Zitrin}}]{rodney18}
{Rodney}, S.~A., {Balestra}, I., {Bradac}, M., {et~al.} 2018, Nature Astronomy, 2, 324, \dodoi{10.1038/s41550-018-0405-4}

\bibitem[{{Sajadian} \& {J{\o}rgensen}(2022)}]{Sajadian_2022}
{Sajadian}, S., \& {J{\o}rgensen}, U.~G. 2022, \aap, 657, A16, \dodoi{10.1051/0004-6361/202141623}

\bibitem[{{Salpeter}(1955)}]{Salpeter_1955}
{Salpeter}, E.~E. 1955, \apj, 121, 161, \dodoi{10.1086/145971}

\bibitem[{{Smith}(2020)}]{Smith_2020}
{Smith}, R.~J. 2020, \araa, 58, 577, \dodoi{10.1146/annurev-astro-032620-020217}

\bibitem[{Sun {et~al.}(2023)Sun, Faucher-Giguère, Hayward, Shen, Wetzel, \& Cochrane}]{sun2023burstystarformationnaturally}
Sun, G., Faucher-Giguère, C.-A., Hayward, C.~C., {et~al.} 2023, Bursty Star Formation Naturally Explains the Abundance of Bright Galaxies at Cosmic Dawn.
\newblock \doarXiv{2307.15305}

\bibitem[{{Trinca} {et~al.}(2024){Trinca}, {Schneider}, {Valiante}, {Graziani}, {Ferrotti}, {Omukai}, \& {Chon}}]{Trinca_2024}
{Trinca}, A., {Schneider}, R., {Valiante}, R., {et~al.} 2024, \mnras, 529, 3563, \dodoi{10.1093/mnras/stae651}

\bibitem[{{Vazdekis} {et~al.}(2016){Vazdekis}, {Koleva}, {Ricciardelli}, {R{\"o}ck}, \& {Falc{\'o}n-Barroso}}]{Vazdekis_2016}
{Vazdekis}, A., {Koleva}, M., {Ricciardelli}, E., {R{\"o}ck}, B., \& {Falc{\'o}n-Barroso}, J. 2016, \mnras, 463, 3409, \dodoi{10.1093/mnras/stw2231}

\bibitem[{Venumadhav {et~al.}(2017)Venumadhav, Dai, \& Miralda-Escud{\'{e} }}]{Venumadhav_2017}
Venumadhav, T., Dai, L., \& Miralda-Escud{\'{e} }, J. 2017, The Astrophysical Journal, 850, 49, \dodoi{10.3847/1538-4357/aa9575}

\bibitem[{Virtanen {et~al.}(2020)Virtanen, Gommers, Oliphant, Haberland, Reddy, Cournapeau, Burovski, Peterson, Weckesser, Bright, {van der Walt}, Brett, Wilson, Millman, Mayorov, Nelson, Jones, Kern, Larson, Carey, Polat, Feng, Moore, {VanderPlas}, Laxalde, Perktold, Cimrman, Henriksen, Quintero, Harris, Archibald, Ribeiro, Pedregosa, {van Mulbregt}, \& {SciPy 1.0 Contributors}}]{scipy}
Virtanen, P., Gommers, R., Oliphant, T.~E., {et~al.} 2020, Nature Methods, 17, 261, \dodoi{10.1038/s41592-019-0686-2}

\bibitem[{{Wang} {et~al.}(2024){Wang}, {Leja}, {Atek}, {Labb{\'e}}, {Li}, {Bezanson}, {Brammer}, {Cutler}, {Dayal}, {Furtak}, {Greene}, {Kokorev}, {Pan}, {Price}, {Suess}, {Weaver}, {Whitaker}, \& {Williams}}]{Wang_2024}
{Wang}, B., {Leja}, J., {Atek}, H., {et~al.} 2024, \apj, 963, 74, \dodoi{10.3847/1538-4357/ad187c}

\bibitem[{{Weisz} {et~al.}(2015){Weisz}, {Johnson}, {Foreman-Mackey}, {Dolphin}, {Beerman}, {Williams}, {Dalcanton}, {Rix}, {Hogg}, {Fouesneau}, {Johnson}, {Bell}, {Boyer}, {Gouliermis}, {Guhathakurta}, {Kalirai}, {Lewis}, {Seth}, \& {Skillman}}]{Weisz_2015}
{Weisz}, D.~R., {Johnson}, L.~C., {Foreman-Mackey}, D., {et~al.} 2015, \apj, 806, 198, \dodoi{10.1088/0004-637X/806/2/198}

\bibitem[{Williams(in prep.)}]{Williams+inprep}
Williams, H. in prep.

\bibitem[{{Williams} {et~al.}(2024){Williams}, {Kelly}, {Treu}, {Amruth}, {Diego}, {Li}, {Meena}, {Zitrin}, {Broadhurst}, \& {Filippenko}}]{Williams+23}
{Williams}, L. L.~R., {Kelly}, P.~L., {Treu}, T., {et~al.} 2024, \apj, 961, 200, \dodoi{10.3847/1538-4357/ad1660}

\bibitem[{Willott {et~al.}(2022)Willott, Doyon, Albert, Brammer, Dixon, Muzic, Ravindranath, Scholz, Abraham, ?tienne Artigau, Brada?, Goudfrooij, Hutchings, Iyer, Jayawardhana, LaMassa, Martis, Meyer, Morishita, Mowla, Muzzin, Noirot, Pacifici, Rowlands, Sarrouh, Sawicki, Taylor, Volk, \& Zabl}]{Willott_2022}
Willott, C.~J., Doyon, R., Albert, L., {et~al.} 2022, Publications of the Astronomical Society of the Pacific, 134, 025002, \dodoi{10.1088/1538-3873/ac5158}

\bibitem[{{Windhorst} {et~al.}(2023){Windhorst}, {Cohen}, {Jansen}, {Summers}, {Tompkins}, {Conselice}, {Driver}, {Yan}, {Coe}, {Frye}, {Grogin}, {Koekemoer}, {Marshall}, {O'Brien}, {Pirzkal}, {Robotham}, {Ryan}, {Willmer}, {Carleton}, {Diego}, {Keel}, {Porto}, {Redshaw}, {Scheller}, {Wilkins}, {Willner}, {Zitrin}, {Adams}, {Austin}, {Arendt}, {Beacom}, {Bhatawdekar}, {Bradley}, {Broadhurst}, {Cheng}, {Civano}, {Dai}, {Dole}, {D'Silva}, {Duncan}, {Fazio}, {Ferrami}, {Ferreira}, {Finkelstein}, {Furtak}, {Gim}, {Griffiths}, {Hammel}, {Harrington}, {Hathi}, {Holwerda}, {Honor}, {Huang}, {Hyun}, {Im}, {Joshi}, {Kamieneski}, {Kelly}, {Larson}, {Li}, {Lim}, {Ma}, {Maksym}, {Manzoni}, {Meena}, {Milam}, {Nonino}, {Pascale}, {Petric}, {Pierel}, {del Carmen Polletta}, {R{\"o}ttgering}, {Rutkowski}, {Smail}, {Straughn}, {Strolger}, {Swirbul}, {Trussler}, {Wang}, {Welch}, {B. Wyithe}, {Yun}, {Zackrisson}, {Zhang}, \& {Zhao}}]{Windhorst+23}
{Windhorst}, R.~A., {Cohen}, S.~H., {Jansen}, R.~A., {et~al.} 2023, \aj, 165, 13, \dodoi{10.3847/1538-3881/aca163}

\bibitem[{{Wirth} {et~al.}(2022){Wirth}, {Kroupa}, {Haas}, {Jerabkova}, {Yan}, \& {{\v{S}}ubr}}]{Wirth_2022}
{Wirth}, H., {Kroupa}, P., {Haas}, J., {et~al.} 2022, \mnras, 516, 3342, \dodoi{10.1093/mnras/stac2424}

\bibitem[{Woodrum {et~al.}(2023)Woodrum, Rieke, Ji, Baker, Bhatawdekar, Bunker, Charlot, Curtis-Lake, Eisenstein, Hainline, Hausen, Helton, Hviding, Johnson, Robertson, Sun, Tacchella, Whitler, Williams, \& Willmer}]{woodrum2023jades}
Woodrum, C., Rieke, M., Ji, Z., {et~al.} 2023, JADES: Using NIRCam Photometry to Investigate the Dependence of Stellar Mass Inferences on the IMF in the Early Universe.
\newblock \doarXiv{2310.18464}

\bibitem[{Yan {et~al.}(2023)Yan, Ma, Sun, Wang, Kelly, Diego, Cohen, Windhorst, Jansen, Grogin, Beacom, Conselice, Driver, Frye, Coe, Marshall, Koekemoer, Willmer, Robotham, D'Silva, Summers, Nonino, Pirzkal, Russell E.~Ryan, au2, Tompkins, Bhatawdekar, Cheng, Zitrin, \& Willner}]{Yan+23}
Yan, H., Ma, Z., Sun, B., {et~al.} 2023, JWST's PEARLS: Transients in the MACS J0416.1-2403 Field.
\newblock \doarXiv{2307.07579}

\bibitem[{Zhang(in prep.)}]{Zhang+inprep}
Zhang, J. in prep.

\end{thebibliography}
\bibliographystyle{aasjournal}



\end{document}